\documentclass[fleqn,usenatbib]{mnras}

\usepackage{newtxtext,newtxmath}
\usepackage[T1]{fontenc}

\DeclareRobustCommand{\VAN}[3]{#2}
\let\VANthebibliography\thebibliography
\def\thebibliography{\DeclareRobustCommand{\VAN}[3]{##3}\VANthebibliography}

\usepackage{graphicx}	% Including figure files
\usepackage{amsmath}	% Advanced maths commands
\usepackage{caption}
\usepackage{titlesec}
\usepackage{arydshln}

\usepackage{comment} %can be removed for submission

\newcommand{\partialdiff}[2]{\frac{\partial #1}{\partial #2}}

\titleformat{\paragraph}
{\normalfont\normalsize\bfseries}{\theparagraph}{1em}{}
\titlespacing*{\paragraph}
{0pt}{3.25ex plus 1ex minus .2ex}{1.5ex plus .2ex}

\title[Detecting giant impacts]{Can giant impacts be directly detected in other star systems?}

\author[P. Tanna et al.]{
Pavan Tanna,$^{1}$\thanks{E-mail: pt426@cam.ac.uk}
Simon Lock,$^{2}$
Amy Bonsor$^{1}$
Simon Hodgkin$^{1}$
Cathie. J. Clarke$^{1}$
\\
$^{1}$Institute of Astronomy, University of Cambridge, Madingley Road, Cambridge CB3 0HA, UK\\
$^{2}$School of Earth Sciences, University of Bristol, Bristol BS8 1RJ, UK\\
}

\date{Accepted XXX. Received YYY; in original form ZZZ}

\pubyear{2026}

\begin{document}
\label{firstpage}
\pagerange{\pageref{firstpage}--\pageref{lastpage}}
\maketitle

\begin{abstract}
% This is a simple template for authors to write new MNRAS papers.
% The abstract should briefly describe the aims, methods, and main results of the paper.
% It should be a single paragraph not more than 250 words (200 words for Letters).
% No references should appear in the abstract.
Giant impacts, collisions between planet-sized bodies, play an important role in planet and moon formation. As we enter a new era of large-scale surveys, such as \textit{Gaia} and LSST at the Vera C. Rubin observatory, there is potential to directly observe the remnants produced in such events and gain insights into the process of planet formation. Here, by modelling the emission and cooling of a series of giant impact remnants, we show that giant impacts are detectable as a sudden brightening followed by gradual dimming in the optical and near-infrared.  Giant impacts between Earth-composition planets were simulated using the smoothed particle hydrodynamics code {\small SWIFT}, with total colliding masses ranging from 0.2 to 4 $M_{\oplus}$. By constraining the location of the photic surface of the post-impact bodies produced by the simulations, the initial luminosities of post-impact bodies were found to be between $5\times10^{-5}$ and $10^{-1}$ solar luminosities, with luminosity falling roughly exponentially on a timescale between 1 and 2000 days. Based on our results, along with estimates for planet and giant impact occurrence rates, we anticipate that between 0 and 14 terrestrial giant impacts will be observed in the full release of \textit{Gaia} epoch photometry in DR4, with at least a comparable number found by LSST. Identifying the remnants of multiple giant impacts will offer a powerful constraint on the frequency of giant impacts in the galaxy and hence the role of such collisions in planet formation.

\end{abstract}

% https://academic.oup.com/DocumentLibrary/mnras/keywords.pdf
\begin{keywords}
planets and satellites: formation -- planets and satellites: detection -- surveys -- techniques: photometric
\end{keywords}

\section{Introduction}

Giant impacts, high-energy collisions between planetary-mass objects that dominate the later stages of terrestrial planet formation, typically occur during the first $\sim$100 Myr of a star's lifetime \citep[e.g.,][]{Quintana2016, Izidoro2017, MacDonald2020, Esteves2022}. During this period, planetary embryos collide with each other and merge to form larger bodies, eventually forming the final configuration of the planetary system \citep{ChambersWetherill1998}. Giant impacts are the most energetic events experienced by terrestrial planets, and can lead to extensive melting and vapourization, alter the composition of planets, and leave post-impact bodies rotating rapidly \citep{LockStewart2017, Nakajima2014, Rufu2018, Carter2020, LOCK2020EPSL, Carter2018, Marcus2009, Marcus2010, Reinhardt2022, Benz2007}. Giant impacts may also control the systems of satellites formed around terrestrial and giant planets \citep{Rosenblatt2020, Kegerreis2018, CameronWard1976, Hartmann1975, Canup2001, Canup2021, CukStewart2012, Lock2018, Nakajima2022} as a fraction of the colliding mass can be injected into orbit around the post-impact body. In fact, the best observational evidence for the occurrence of giant impacts is the existence of our own Moon, which probably formed in the aftermath of an impact between the proto-Earth and another planetary body (often known as Theia) \citep{CameronWard1976, Hartmann1975}. In our solar system, giant impacts also likely played a role in the formation of the Martian moons \citep{Craddock2011, Rosenblatt2012, Citron2015, Canup2018, Rosenblatt2020}, Mercury's large iron core \citep{Benz2007,Asphaug2014}, the Pluto-Charon system \citep{Canup2021, Denton2025}, and Uranus' obliquity \citep{Korycansky1990, Kegerreis2018}. 

The ability of giant impacts to shape planets and planetary systems makes them a critical piece in understanding planet formation. As the type example of a planetary collision, a large number of studies have been conducted in the context of the formation of our Moon. To meet the increasingly precise observational constraints \citep{Lock2020SSR}, a large range of impact scenarios have been proposed \citep{Canup2001, Reufer2012, CukStewart2012, Canup2012, Kegerreis2022, Rufu2018, Lock2018} from a comparatively low energy collision between the proto-Earth and a Mars-mass impactor, often referred to as the `canonical Moon-forming impact' \citep{Canup2001}, to impacts between more equal mass bodies and/or at high velocities \citep{CukStewart2012,Canup2012,Lock2018} for which the interacting specific energy can be more than an order of magnitude greater \citep{LockStewart2017,Carter2020}. 

\begin{comment}
The original idea was proposed by \cite{Hartmann1975} and \cite{CameronWard1976} to explain the angular momentum of the Earth-Moon system, the Moon's relatively small iron core, and the inferred high temperature of the early Moon. Initial numerical studies found that following a grazing impact between the proto-Earth and a Mars-mass impactor, a large amount of debris accreted to form a moon, potentially matching these constraints \cite{canup2001}. This particular impact scenario is often referred to as the `canonical Moon-forming impact'. 

Additional and improved observational constraints on the formation of the Moon \citep{Lock2020SSR} have called the viability of the canonical scenario into question and led to a variety of different types of giant impacts to be proposed and investigated. These scenarios include: higher velocity `hit and run' impacts \citep{Reufer2012} (with excess angular momentum and debris ejected from the system); higher angular momentum impacts, for example impacts where the target is spinning very rapidly \citep{CukStewart2012} or where the two colliding bodies have a roughly equal mass \citep{Canup2012}, (where the excess angular momentum is transferred away from the Earth-Moon system by orbital interactions \citep{CukStewart2012, Wisdom2015, Cuk2016, Cuk2020}); \cite{Kegerreis2022} suggest that a moon can form from a single impact in the period of a few days; and \cite{Rufu2018} suggest that the Moon may have formed from a series of smaller moonlets merging together. 
\end{comment}

The post-impact bodies produced by this range of proposed scenarios are typically very hot, substantially melted and vapourized, oblate, and extend to many times the size of the bodies that collided to produce them. In particular, higher angular momentum impacts can produce `synestias': rapidly rotating structures where the post-impact disk and a co-rotating central region become a single continuous structure \citep{LockStewart2017}. First-order calculations suggest that post-impact bodies could survive in an inflated, significantly-vapourised state for around $10^{2}$ to $10^{3}$ yr following an impact \citep{Lock2018, LOCK2020EPSL}. 

\begin{comment}
Smoothed particle hydrodynamics (SPH) is the most commonly used technique to model giant impacts. As it is a Lagrangian, particle based method, it maintains a high resolution in regions where the mass is concentrated, as opposed to grid based methods, which have a fixed resolution across the whole space of the simulation. Particle based codes also can be easily combined with gravity solvers, allowing self gravity, a key effect in giant impacts, to be easily modelled. 
\end{comment}

Recently, signatures of giant impacts have been identified in other systems -- primarily discovered and observed through unusual excess and/or variability in the emission from their host system (separate from normal stellar variability) -- opening an exciting new avenue to observationally constrain the frequency and outcome of giant impacts \citep{Kenworthy2023, Su2022, Su2019, Lisse2009, Schneiderman2021}. The observed variability is thought to be caused by sudden increases in luminosity from the afterglow of the post-impact body, emission from the large amounts of hot or cold dust and gas ejected into the star system, and/or later (often irregular) transits of the debris across the line of sight between Earth and the host star. The post-impact bodies themselves are theorised to have temperatures ranging from a few hundred to thousands of Kelvin depending on composition \citep{Lock2018, Kenworthy2023} with their emission mostly in the near-infrared and optical, while the ejected dust is passively heated by the star and is often colder (tens to hundreds of Kelvin) and emits in the mid-infrared. Two particularly notable cases are the variability observed in the HD 166191 system by \cite{Su2022}, with warm dust ejected from giant impact proposed as the most likely explanation, and the ASASSN-21qj system where brightening of the system in the infrared is consistent with direct emission from a post-impact body produced by a collision involving planets of several to tens of Earth masses, with a later irregular transit caused by impact debris \cite{Kenworthy2023}. These individual discoveries suggest that a systematic search using large scale astronomical surveys may discover evidence of many more impacts.

In this new era of astronomical surveys, with facilities such as the Zwicky Transient Facility (ZTF), \textit{Gaia} and the Legacy Survey of Space and Time (LSST) at the Vera C. Rubin Observatory, we have a unique opportunity to conduct an extensive search for the sudden increase in luminosity from planetary systems following giant impacts. The latest data release of the \textit{Gaia} catalogue (DR4) will contain nearly 2 billion sources with photometric data over 5 years, with an average cadence of 30 days \citep{GaiaReleaseInfo}.  LSST will accurately measure around 10 billion stars in the southern sky with millimagnitude accuracy \citep{LSSTBook}, extending \textit{Gaia}'s accuracy down to magnitudes of around 27, and improving its cadence down to around 3 days. With such an unprecedented number of sources, even rare transient events such as giant impacts may appear multiple times in the combined dataset. 

In this paper, we aim to determine the increase in luminosity of a planetary system following a giant impact between terrestrial planets, in order to assess the feasibility of detecting giant impacts with current and upcoming surveys. By simulating a range of planetary impactors using the smoothed particle hydrodynamics (SPH) code {\small SWIFT} and estimating the luminosity of the post-impact body as it cools, we calculated the initial luminosity and light curves of a variety of post-impact bodies from a few days after first contact, as described in \S\ref{sec:methods}. The results of these calculations and comparisons with the capabilities of various photometric surveys are detailed in \S\ref{sec:results}. The potential limitations of the analysis and the implications of the results are discussed in \S \ref{sec:discussions} and we conclude in \S\ref{sec:conclusions}. In this work we only consider giant impacts between rocky bodies below 2 $M_{\oplus}$, and hence on a somewhat similar scale to Moon-forming impacts. Henceforth, unless otherwise specified, references to giant impacts includes only this scale of impacts.

\section{Methods}\label{sec:methods}

We assess the observability of a range of giant impacts based on the luminosity and long-term evolution of the post-impact structure they produce. Note that here, for simplicity, we are calculating only the emission from the gravitationally bound largest remnant and not emission from unbound material that is injected onto separate orbits around the host star. The intrinsic luminosity of hot unbound material could have dominated the emission over the duration of the initial impact (tens of hours), partly as the impact remnant could have been obscured by the unbound material, but the unbound material expands and cools quickly reducing its intrinsic luminosity and ability to obscure the post-impact body. As seen in several systems \citep[e.g.][]{Su2019}, passive heating of the unbound material by the host star can lead to an observable infrared excess on longer timescales. However, we are focusing here on optical surveys and so only the closest-in debris disks will contribute significantly to the observed emission at relevant wavelengths. We will consider the effect of the unbound material on the observability of giant impacts in Section~\ref{sec:discussion:cooling} and an upcoming study. 

The structure of several example post-impact bodies were determined by simulating giant impacts using the Smoothed Particle Hydrodynamics (SPH) code {\small SWIFT} \citep{SWIFT}, as described in \S\ref{sec:SPH}. To calculate the luminosity of a post-impact body from the SPH output, it is necessary to resolve the structure to very low pressures ($<\sim10^{5}$ Pa) where the structure becomes optically thin. However, the resolution required to resolve such low pressures is not computationally feasible. Thus, we developed a method to extrapolate beyond the resolved regions of the SPH structure to the outer, optically-thin regions as described in \S\ref{sec:extrapolation}. The initial luminosity (\S\ref{sec:photosphere}) and cooling (\S\ref{sec:cooling}) of the post-impact body can then be calculated from the extrapolated structure.

\subsection{Smoothed particle hydrodynamics impact simulations}\label{sec:SPH}

18 giant impacts were simulated using {\small SWIFT}\footnote{Version 0.9.0}, with impact parameters shown in Table~\ref{tab:simulation_params}. This impact suite samples only a small subset of the large variety of potential impacts that happen across different planetary systems and our simulation set is by no means comprehensive. All the impacts were variants on a single example impact \citep[similar to the Moon-formation scenario proposed by][]{Canup2012} with either one or two initial parameters modified to investigate how this affects the resulting post-impact bodies. We did not vary impact velocity, instead taking an impact velocity of 1.1 times the mutual escape velocity -- corresponding to the mode in terrestrial planet formation models \citep{Raymond2009, CukStewart2012, Quintana2016}. Higher-velocity impacts generally produce more extended post-impact bodies and so would be expected to be more easily observable, but a detailed examination of the effect of impact velocity is left to future work.

The initial conditions for the colliding planets were generated using the {\small WoMa}\footnote{Version 1.2.0} package \citep{woma}. {\small WoMa} generates an arrangement of SPH particles based on a set of initial conditions and properties of the impacting planets and a specified number of particles of equal mass. The planets generated have approximately $10^5$ particles, with variation due to the SEAGen algorithm \citep{Kegerreis2019} used by {\small WoMa} .

The cores and mantles of the colliding planets were modelled as pure iron and forsterite respectively, using the ANEOS-2020 pure iron and ANEOS-2019 forsterite equations of state \citep{ANEOS2020, ANEOS2019} (which were also used during the impact simulation). The core mass was set at 33\% of the total planet mass to be the same as present-day Earth. The core entropy was set to 1750 J/K (corresponding to a temperature of 4000 K at Earth's core mantle boundary pressure) and the mantle entropy was set to 3027 J/K (corresponding to a 1 bar mantle temperature of 2150 K, i.e., a molten surface). 

The {\small SWIFT} simulations were initiated 30 minutes before impact to allow for tidal deformation of the colliding bodies, and were run for a total simulation time of 40 hours, which was sufficient time for the post-impact bodies to reach a pseudo steady state. {\small SWIFT} accurately traces gravity and pressure forces, but does not include cooling or phase separation, so subsequent evolution was calculated using a different approach.  For computational efficiency, {\small SWIFT} enforces a maximum smoothing length for the calculations, beyond which particles reach a fixed minimum density and behave ballistically.  The maximum smoothing length used here was 1.2 $R_{\oplus}$, corresponding to a minimum density of $0.17\,\mathrm{kg}\,\mathrm{m}^{-3}$ to $1.4\,\mathrm{kg}\,\mathrm{m}^{-3}$ (depending on the total mass of the particles in the simulation), which ensured that the density floor was not reached until around 10~$R_{\oplus}$ from the centre of the post-impact bodies.  The impactor and target planets were not settled prior to being injected into the impact simulation.

\begin{table}
    \centering
    \begin{tabular}{cc|ccccc}
        \multicolumn{2}{c|}{Simulation} & \multicolumn{5}{c}{Initial Parameters}\\
        Set & \# & No.\ of particles & $M_{\mathrm{total}}$ / $M_{\oplus}$ & $\gamma$ & $b$ & $v_{\mathrm{imp}}$ / $v_{\mathrm{esc}}$ \\ \hline
        A & 0 & 104226 & 1.00 & 0.50 & 0.50 & 1.1 \\
         & 1 & 112252 & 0.20 & 0.50 & 0.50 & 1.1 \\
         & 2 & 106160 & 0.50 & 0.50 & 0.50 & 1.1 \\
         & 3 & 114000 & 0.75 & 0.50 & 0.50 & 1.1 \\
         & 4 & 99738 & 1.50 & 0.50 & 0.50 & 1.1 \\
         & 5 & 112312 & 2.00 & 0.50 & 0.50 & 1.1 \\
         & 6 & 99152 & 3.00 & 0.50 & 0.50 & 1.1 \\
         & 7 & 103290 & 4.00 & 0.50 & 0.50 & 1.1 \\
        B & 8 & 111197 & 0.52 & 0.05 & 0.50 & 1.1 \\
         & 9 & 109247 & 0.60 & 0.17 & 0.50 & 1.1 \\
        C & 10 & 109092 & 0.75 & 0.33 & 0.50 & 1.1 \\
         & 11 & 107820 & 0.15 & 0.33 & 0.50 & 1.1 \\
         & 12 & 106744 & 0.37 & 0.33 & 0.50 & 1.1 \\
         & 13 & 103444 & 1.50 & 0.33 & 0.50 & 1.1 \\
         & 14 & 114678 & 3.00 & 0.33 & 0.50 & 1.1 \\
        D & 15 & 104226 & 1.00 & 0.50 & 0.10 & 1.1 \\
         & 16 & 104226 & 1.00 & 0.50 & 0.20 & 1.1 \\
         & 17 & 104226 & 1.00 & 0.50 & 0.30 & 1.1 \\
         & 18 & 104226 & 1.00 & 0.50 & 0.40 & 1.1 \\
         & 19 & 104226 & 1.00 & 0.50 & 0.60 & 1.1 \\
         & 20 & 104226 & 1.00 & 0.50 & 0.70 & 1.1 \\
        E & 21 & 112252 & 0.20 & 0.50 & 0.10 & 1.1 \\
         & 22 & 106160 & 0.50 & 0.50 & 0.10 & 1.1 \\
         & 23 & 112312 & 2.00 & 0.50 & 0.10 & 1.1 \\
         & 24 & 103290 & 4.00 & 0.50 & 0.10 & 1.1 \\
    \end{tabular}
    \caption{This table presents the initial parameters chosen for the simulations described in \S\ref{sec:SPH}, including the total mass involved in the impact ($M_{\mathrm{total}}$), the impact parameter ($b$), the impactor-to-total mass ratio ($\gamma$) and the impact velocity relative to the mutual escape velocity ($v_{\mathrm{imp}}/v_{\mathrm{esc}}$). Simulations are divided into different sets depending on their impact parameters relative to our reference simulation (0). Set A has a changing $M_{\mathrm{total}}$, set B has a changing $b$, set C has changing $M_{\mathrm{total}}$ with a lower value of $b$, set D has changing $M_{\mathrm{total}}$ with a lower value of $\gamma$, and set E has different total masses with a smaller value of $b$.}
    \label{tab:simulation_params}
\end{table}

\subsection{Determining the properties of the post-impact structure} \label{sec:extrapolation}

The outer regions of the post-impact bodies generated in our simulations are predominantly composed of silicate vapour and condensates (potentially both solid and liquid), which becomes more opaque further into the structure. Estimating the thermodynamic properties within this region is crucial for calculating the opacity of the structure, estimating the depth to which the interior structure of the impact is visible from outside the body, and hence determining the emission from the body. 

However, the structure of the post-impact body, and the multiphase, multicomponent processes governing its evolution (see below) are highly complex and not well understood \citep{Lock2017, Lock2018, LOCK2020EPSL}. The aim of this study is to perform the first estimate of the emission of a post-impact body and so we take a simplified approach, approximating the post-impact body as having a pseudo-1D structure based on the midplane structure. 

The raw simulation output contains the thermal properties (e.g., pressure, entropy, temperature) and position of each SPH particle. To calculate the thermal properties at a given point using the SPH kernel function we used the {\small SWIFTsimIO} package \citep{swiftsimio}. As the post-impact body is axisymmetric, we averaged the thermal properties (density, temperature, pressure and specific entropy) around the rotational axis by sampling points on a circle in the equatorial plane and taking the mean of those points. We note that the radial profiles of density, temperature, pressure and specific entropy are each averaged independently, and the averaged quantities do not therefore constitute a strictly self-consistent thermodynamic state: re-evaluating the equation of state at the mean density and temperature would not in general return the mean specific entropy. Because we use specific entropy only to identify the isentrope used for the extrapolation, and density and temperature only for the optical-depth and emission calculations, this inconsistency does not propagate into our results.

This was done at different radii, giving us the thermal properties in the equatorial plane of the post impact body in radial bins. We set 400 linearly spaced radial bins between the post-impact body's centre of mass and the maximum radius, which was the minimum of either the hill sphere (described in \S\ref{sec:hill}) or 10 $R_{\oplus}$.

Resolving the optically thin layers of the post-impact bodies directly in SPH requires computationally unfeasibly high particle resolutions. Further, a simultaneous strength and key limitation of {\small SWIFT} simulations is that once particles have reached the maximum smoothing length, and the density floor has been reached, pressure forces are no longer accurately simulated as the particles are not interacting with each other. Thermodynamic properties calculated in these regions are, by definition, inaccurate. In our simulations, minimum density particles are typically found beyond around 10 $R_{\oplus}$ from the centre of the post-impact body. We therefore take the approach to extrapolate the thermodynamic properties of the post-impact body outwards, from the resolved interior regions, adding additional radial bins until we reach the edge of the Hill sphere. 

To do this, we assume that the post-impact body is axisymmetric, in rotating hydrostatic equilibrium, and that its gravitational field can be approximated as a point mass (as the vast majority of the mass is concentrated close to its centre). The pressure of the fluid in the equator of the post-impact body can thus be found by integrating outwards from the last radial bin using the following equation:

\begin{equation}
    \partialdiff{P}{r} = -\rho \left( - \frac{GM}{r^{2}} + r\Omega^{2}\right) ,
    \label{eq:extrapolation}
\end{equation}

where $r$ is the distance from the centre of mass, $P$ and $\rho$ are the fluid's pressure and density, $G$ is the gravitational constant, $M$ is the total mass of the post-impact body, and $\Omega$ is the fluid angular velocity which is assumed to be a function of distance from the rotation axis only. This equation is solved numerically (using {\small SciPy} \citep{scipy}) given a description of the density and angular velocity as a function of pressure and/or position.

We estimated a radial angular velocity profile in the equatorial plane by analysing the angular velocity of the SPH particles from each simulation. Beyond the co-rotation point (identified with least-squares curve fitting), interior to which all the particles move together with the same angular velocity, the angular velocity follows an approximate power law, with an angular velocity slightly below the Keplerian orbital velocity due to the additional pressure forces from the silicate fluid that support against gravity \citep{LockStewart2017}. The power law was fitted to particles within 0.1 $R_{\oplus}$ of the midplane of the post-impact disk. A separate power law was calculated for each individual simulation. We assume that the outer regions of the post-impact body are approximately isentropic, as has been observed in previous SPH giant impact studies \citep{LockStewart2017, Nakajima2014}, taking the constant entropy value as the entropy of the last radial bin from which the extrapolation begins.

We find the density and temperature of the silicate fluid from the pressure and the entropy using the forsterite equation of state \citep{ANEOS2019}. 

\subsubsection{Accounting for droplet physics}\label{sec:droplet_physics}

The presence and continued condensation of droplets in the post-impact body will affect its structure and optical depth. Current SPH codes do not model separation of vapour and liquid. Droplets condensing in the vapour, either due to decompression or radiative cooling, are estimated to grow to on the order of millimetres in size and decouple from the gas \citep{Lock2018}. Given the vapour velocity is slightly lower than Keplerian velocity outside the corotating region, the droplets (which are not significantly supported by the pressure gradient in the fluid) experience a significant headwind. The resulting drag causes droplets to fall towards the higher density, hotter central regions of the body, potentially revapourising. We hence developed a method to account for infalling droplets during our luminosity and cooling calculations.

Drag forces dominate the infall of droplets. Using the formula for aerodynamic drag, the infall timescale for droplets decoupling from the vapour can be estimated as

\begin{equation}
    t_{\mathrm{infall}} = \frac{m_{\mathrm{drop}} v_{\mathrm{drop}}}{F_{\mathrm{drag}}} = \frac{2 \rho_{\mathrm{drop}} D_{0} v_{\mathrm{drop}}}{\rho_{\mathrm{vap}} C_{D} v_{\mathrm{rel}}^{2}} , 
    \label{eq:infall_time}
\end{equation}

where $F_{\mathrm{drag}}$ is the drag force, $v_{\mathrm{rel}}$ is the difference between the droplet velocity (assumed to be the velocity for a circular Keplerian velocity) and the vapour velocity (calculated from the power law described previously), $v_{\mathrm{drop}}$ is the (circular Keplerian) orbital velocity of the droplets, $C_{D}$ is the drag coefficient, \citep[assumed to be 0.5, the value for a sphere in the high Reynolds regime][]{Beard1976}, $D_{0}$ is the droplet diameter, $m_{\mathrm{drop}}$ is the droplet mass, $\rho_{\mathrm{vap}}$ and $\rho_{\mathrm{drop}}$ are the densities of the vapour and droplets respectively.

When estimating the cooling of, and emission from, post-impact bodies, if the infall timescale is sufficiently short in a given radial bin, droplets are removed by altering the thermodynamic properties of the material in that bin. The timescale threshold was chosen to be $10^{4}$ seconds, typical of the orbital timescale in the post impact body. After the initial extrapolation of thermodynamic properties performed assuming constant entropy, any radial bin with condensing and infalling droplets (i.e. the pressure and entropy is inside the vapour dome) has its entropy set to that at the vapour edge of the liquid-vapour phase boundary to account for the droplet infall and its pressure is kept constant. Density and temperature are then recalculated from these pressure and entropy values. This technique is similar to previous giant impact studies such as \cite{Lock2018}. Here, we do not consider the potential for re-vapourization of the infalling condensates.

%An example of this extrapolation is demonstrated on a phase diagram in Fig.~\ref{fig:phase_diagram}, which shows the thermal profile after the constant entropy extrapolation and the following droplet removal.

\subsection{Determination of the photosphere} \label{sec:photosphere}

To calculate the luminosity of a post-impact body, the area and temperature of its photosphere must be estimated to find the emission at this surface. Here, we utilise the theoretical model for the wavelength-integrated absorption coefficient of a silica fluid \citep{Kraus2012}, which accounts for the absorption caused by the silicate vapour (based on a Drude semiconductor model) and the condensed droplets (assuming perfect scattering):

\begin{align}
    \alpha &= \frac{6}{D_{0}} \frac{V_{\mathrm{liq}}}{V_{\mathrm{avg}}} + B_{0} \left( \frac{\rho}{\rho_{n}}\right)^{1/3} \left(\frac{T}{T_{n}}\right)e^{-B_{1}\frac{T_{n}}{T}} e^{-B_{2}\frac{\rho T_{n}}{\rho_{n} T}} \label{eq:krauss} \\
    B_{0} &= 6 \times 10^{11} \, \mathrm{m} \nonumber \\
    B_{1} &= 37 \nonumber \\
    B_{2} &= -11.6 \nonumber \\
    \rho_{n} &= 1900 \, \mathrm{kg} \, \mathrm{m}^{-3} \nonumber \\
    T_{n} &= 4150 \, \mathrm{K} \nonumber
\end{align}

where $D_{0}$ is the diameter of the droplets, $\frac{V_{\mathrm{liq}}}{V_{\mathrm{avg}}}$ is the liquid (condensate) volume fraction, $T$ is the temperature, and $\rho$ is the density of the vapour. $B_i$ are parameters that were fit by \cite{Kraus2012} to density functional theory calculations. \cite{Lock2018} estimated the droplet size to be around $10^{-3}$~m and we adopt this value throughout this work.  

%which has been used with corrections in previous giant impact studies such as \cite{Lock2018}
%The absorption of the fluid is plotted on the phase diagram in Fig.~\ref{fig:phase_diagram}, which shows the silicate fluid becoming more optically thin at lower pressures in the vapour phase.

Using the thermodynamic properties in the equator of the outer regions of the post-impact body (as found from the extrapolation model), the absorption ($\alpha$) can be integrated from outside the structure with radial distance ($r$) inwards to find the optical depth of the structure ($\tau$). 

\begin{equation}
    %\int_{r_{\text{photosphere}}}^{\infty} \alpha(r) dr = \frac{2}{3}
    \tau = \int_{r}^{\infty} \alpha(r) dr .
    \label{eq:photosphere}
\end{equation}

We define the photosphere as the point where $\tau = 1$. This is the maximum depth from which radiation can directly escape from the post-impact body to be observed. The luminosity is calculated using Stefan-Boltzmann law with the radius and the temperature at the photosphere surface in the midplane, with the assumption that the photosphere is spherical.

\subsection{Effect of planetary orbital parameters} \label{sec:hill}

The SPH simulations do not simulate any external gravitational fields and thus do not take into account the fact that material that would be bound to the post-impact body in isolation, could actually be more strongly bound to the host star and eventually lost from the post-impact body. The Hill sphere allows us to estimate how much material remains bound by imposing a maximum size for the post-impact body. We make the assumption that any material further from the centre of the post-impact body than the Hill radius becomes unbound to the post-impact body and orbits the host star instead. This unbound material will cool quickly and likely show up as warm dust surrounding the star, potentially visible in the mid infrared. We therefore begin the integration to calculate the optical depth (Eq.~\ref{eq:photosphere}) at the Hill radius and assume that there is no material beyond this radius.

The size of the Hill sphere entirely depends on the planetary system architecture. A more massive post-impact body, a greater distance of the body from the star or a less massive host star will make the Hill sphere larger. Hence, the photosphere and cooling calculations have an imposed maximum size set by this Hill sphere. A smaller Hill sphere could lead to a lower luminosity, due to the reduced area for emission. The Hill sphere can be very small, potentially around 15 $R_{\oplus}$ in compact systems like TRAPPIST-1, hence this could have quite a large effect on our results, and so we consider multiple values for the Hill sphere in this work.

We calculate the Hill radius of the post-impact body as:

\begin{equation}
    R_{H} = \sqrt[3]{\frac{GMT_{\mathrm{orb}}^{2}}{12\pi^2}} , \label{eq:hill2}
\end{equation}

where $T_{\mathrm{orb}}$ is the body's orbital period around its host star, $M$ is the mass of the post-impact body, and $G$ is the gravitational constant.

Given the mass of the post-impact structure is known from our SPH simulations, the only free parameter for the Hill radius is the orbital period of the post-impact structure (our SPH simulations do not account for the orbits of the colliding planets). We analyse each simulation three times with orbital periods of 1, 10 and 100 days to cover the range of orbital periods likely for planets around low mass stars \citep{Neil2020}. 

\subsection{Cooling of the post-impact structure} \label{sec:cooling}

In order to detect the emission from the aftermath of impacts, it is crucial to consider the long-term evolution of post-impact bodies. The impact remnant must persist for a sufficient duration to allow the detection of any increase in brightness, and the brightness evolution might be able to distinguish a post-impact body from other transient events. The cooling of post-impact bodies occurs on timescales of hundreds to thousands of years \citep{LOCK2020EPSL}, significantly longer than the 40~hour SPH simulation time and an additional cooling model is required.

Here, we aim to determine a reasonable lower bound on the timescale for cooling, and hence the detectable lifetime, of post-impact bodies. It is difficult to analytically calculate the heat transport (either with convection, rain-out of droplets, or radiation) within the post-impact body \citep[see discussion in][]{Lock2018,Lock2020SSR}. We hence build on our post-impact structure model to produce a simple cooling model, making assumptions that promote the determination of the lower limit on cooling timescale. 

The total energy lost from the post-impact structure at each time step is set by the area and temperature of the photosphere using Stefan-Boltzmann's law. This loss of energy is compensated by the internal energy of a fraction of the post-impact body in radial bins at pressures below a set threshold being reduced, with the internal energy of all relevant bins being reduced by the same fraction at each time step. The use of a pressure threshold means that only the energy in the outer regions of the post-impact body is available to be radiated away. The default value we adopt is $10^9$~Pa, which is typical of the initial pressure of the transition between the corotating inner region and disk-like region in post-impact bodies \citep{LockStewart2017} and roughly an order of magnitude greater than the critical point of silicates \citep{Kraus2012,Caracas2023,Davies2020,Xiao2018}. This method produces a lower bound estimate of the cooling time as it assumes that heat transport to the photosphere is highly efficient. The effect of varying the pressure threshold is examined in detail in the discussion (Section~\ref{sec:discussion:cooling}). 
After each cooling step, the thermal properties were recalculated using the forsterite equation of state assuming the density is constant and internal energy reduced by the amount lost through radiation. Redistribution of mass to reach pressure equilibrium in the post-impact structure was not considered. 

To account for the cooling of material outside the photosphere, the vapour outside the photosphere was cooled proportionally at the same rate as the post-impact body within the photosphere. In each timestep, the material outside the surface of the photosphere loses the same fraction of its energy as the material inside the photosphere. The cooling was evolved with a fixed timestep over the total evolution period, with the calculation terminating after 100 years.

The energy loss from radiation causes the vapour to cool and condense into droplets which fall inwards. On each timestep, the droplets are removed as described in \S\ref{sec:droplet_physics} and the location of the photosphere is recalculated. Redistribution of energy, mass, and angular momentum by falling condensates was not considered, which again leads to an underestimate in the cooling timescale \citep{Lock2018}. During cooling, the outer regions of the post-impact body thus shrink and lose mass over time.

%The most accurate technique to calculate the cooling would be to apply the radiative transfer equation to the system. This is mathematically and computationally complex as the opacity of the structure changes significantly as the structure cools. We can assume all radiation originates from the surface of the photosphere as a simplification to make the cooling estimation possible. This provides a luminosity and a region within to cool. However, this does not account for the radiation and cooling of the vapour outside the photosphere. The photosphere is an average last scattering surface for the radiation leaving the structure, in reality, radiation originates from a continuous region of the post-impact body.

\subsection{Estimating observability}

With our method to estimate the luminosity and cooling of a post-impact body, we can begin to estimate the likelihood of observing one. There are two parts to this estimate. Firstly, we need to compare the range of luminosities and cooling times from our impact simulations with the capabilities of the \textit{Gaia} and LSST surveys to estimate if these surveys are capable of detecting post-impact bodies (\S\ref{sec:dectability}). Secondly, we need to estimate the frequency of giant impacts in a population of stars from our current knowledge of planet formation and planet occurrence rates (\S\ref{sec:frequency}). With these estimates, we can then predict the number of impacts that we expect to observe with \textit{Gaia} and LSST using a Monte Carlo method. Starting with the number of stars expected in each survey, we randomly select stars younger than 100 Myr, young enough to have a high probability of hosting giant impacts. For each of those stars, we randomly assign a number of rocky planets and the number of impacts each planet had during its formation, based on our current best estimates of planet occurrence rates and giant impacts. With these values we can calculate an impact rate for each star systems. Using a Poisson distribution parametrized by this impact rate and a set observing time, we then calculate the number of impacts we expect to happen around these stars during the duration of observations.

Given our limited knowledge of post-impact body luminosity and our small number of simulations, we take a triangle distribution as a good first order estimate for the distribution of the luminosities of post-impact bodies. The lowest, highest and mean luminosity of our range of simulated impacts were taken as the minimum, maximum and peak of the distribution. Using this distribution we then randomly assign a luminosity to the impacts in our Monte Carlo analysis and determine how many of these impacts will be observable given the sensitivity of \textit{Gaia} and LSST. We perform our Monte Carlo analysis 300 times to produce a range for the number of impacts we expect to observe in a five and ten year periods for both \textit{Gaia} and LSST.

\subsubsection{Detectability of post-impact bodies} \label{sec:dectability}

The potential for giant impacts to show up as detectable changes in the brightness of combined star and planet systems in stellar surveys is dependent on the luminosity and cooling timescales of the post-impact structures.  To detect the change in luminosity from a stellar system due to a giant impact, the change in flux, or equivalently apparent magnitude, detected from the impact must be greater than the error on the measurements and this change must last longer than the survey's cadence. We hence define a detection as when the change in luminosity caused by an impact is 3 times the uncertainty of the magnitude measurement for a 3-sigma detection. We define the cooling timescale as the time for the luminosity for the post-impact body to decrease to half its initial luminosity and use this to verify if the impacts modelled last long enough to be detected.

Our primary focus is on the \textit{Gaia} survey, given its photometric sensitivity and its present data releases provide a comprehensive view of the stars within the catalogue. \textit{Gaia} DR4, which is due to be released in December 2026, will contain epoch photometry for all sources for a 5 year period \citep{GaiaDR4Web}. This epoch photometry has the potential to reveal giant impacts, if they can be disentangled from other signatures. 

\paragraph*{\textit{Gaia}} \label{sec:gaia}

\begin{figure}
    \centering
    \includegraphics[width=\linewidth]{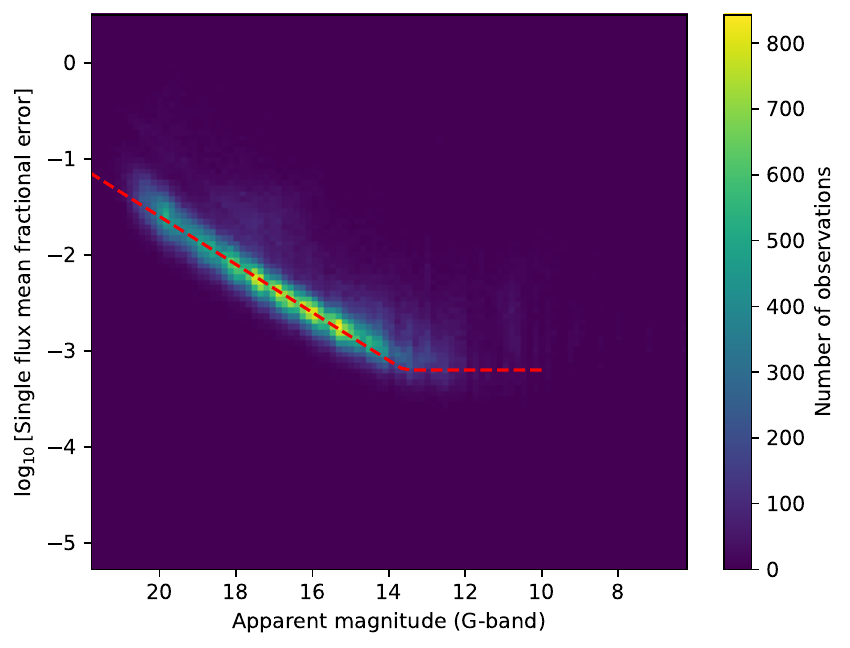}
    \caption{2D histogram showing the error on a single flux measurement of \textit{Gaia} against the apparent magnitude (G-band). The red line shows the fitted relationship between the two values that is used to estimate the detectability of post-impact bodies.}\label{fig:single_flux_error}
\end{figure}

Using epoch photometry data \footnote{The epoch photometry is the collection of all the flux measurements across the duration of the \textit{Gaia} mission.} from \textit{Gaia} DR3, we estimated a relationship between a source's apparent magnitude and the fractional error on a single flux measurement of that source. This estimated relationship is shown in Fig.~\ref{fig:single_flux_error}. This provides the uncertainty of the magnitude used to verify if an impact is observable. 

Using this relationship, we estimated the error on the individual flux measurements for all stars in the \textit{Gaia} catalogue \footnote{Due to the large size of \textit{Gaia} DR3, the analysis was performed on only a random sub-sample of around 2 million stars to reduce computation time.} and compared them to the estimated change in flux from a giant impact. 

We calculate the new g-band magnitude of a star system (of magnitude $m_{g,*}$) after an impact with a brightness of $n$ times the error on $m_{g,*}$ ($\Delta m$) with the below formula:

\begin{equation}
    m_{g,i} = m_{g,*} - \log\left( 10^{\frac{n\Delta m}{2.5}} - 1\right) . 
\end{equation}

\begin{figure}
    \centering
    \includegraphics[width=\linewidth]{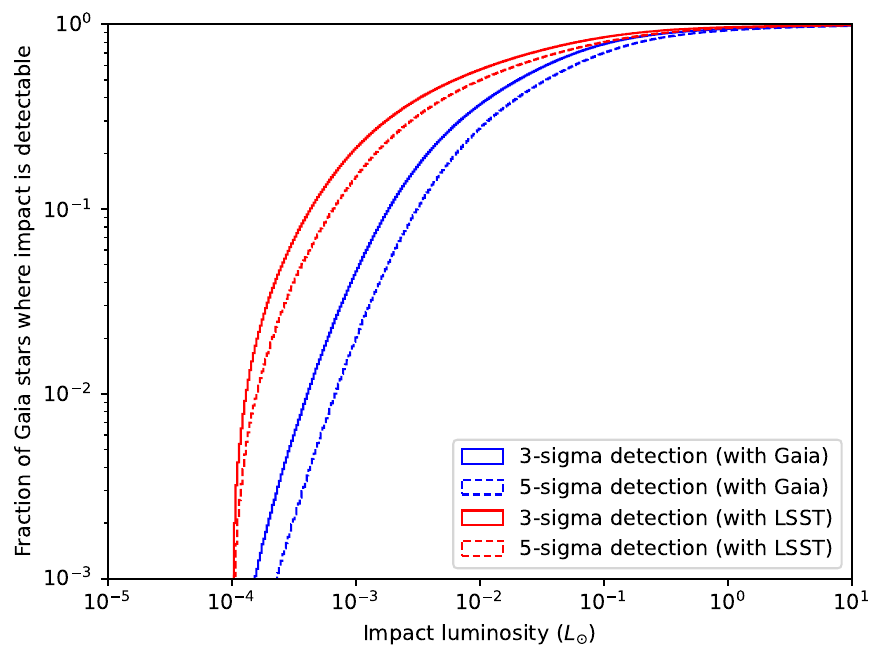}
    \caption{Cumulative histogram showing, for a given impact luminosity, the proportion of stars in the \textit{Gaia} sample where this impact is detectable, as calculated in \S\ref{sec:gaia}. The plot shows this detectable proportion for \textit{Gaia}'s and LSST's sensitivity with 3 and 5 sigma detections. This plot does not take into account that giant impacts only happen around young stars.}\label{fig:cumulative_detectability}
\end{figure}

We use a bolometric correction factor \citep{Creevey2023} to convert this brightness to solar luminosity, allowing us to calculate the luminosity of the new source around the star. We compare this to the post-impact structure calculated by our model. We find the relationship between the new source luminosity (corresponding to the luminosity of a newly created post-impact structure) and the proportion of stars in \textit{Gaia} where this new source would be detectable. This relationship is shown in Fig.~\ref{fig:cumulative_detectability} .

\paragraph*{LSST}

The Vera C Rubin observatory will survey the sky at unprecedented depth, producing the Legacy Survey of Space and Time \citep{LSST} (LSST), providing another dataset in which to search for giant impacts. LSST is expected to accurately measure the properties of 10 billion stars across half the sky. For apparent magnitudes between 17 and 20, LSST will have a higher precision than \textit{Gaia}, with an estimated precision for a single observation of 5 millimagnitudes \citep{LSST}. Furthermore, LSST will be able to detect stars up to a magnitude 24 on a single observation, albeit with much less precision.

As first light of the Rubin observatory's Simonyi Survey Telescope was only recently completed in 2025, the capabilities of LSST are much less well known than \textit{Gaia}. The alert stream starts this year and LSST Data Release 1 is expected in 2028 \citep{RubinPlans}, with transient detection demonstrated by \cite{Dong2025} with Data Preview 1.

We produce two estimates for the expected number of giant impacts LSST will detect. One is a very conservative estimate, assuming LSST only observes stars already detected by \textit{Gaia}. However, given LSST's higher sensitivity, we would expect a moderate increase in the expected number of impacts (see Fig.~\ref{fig:cumulative_detectability}). 

LSST should be able to detect many dim M stars that \textit{Gaia} is not sensitive enough to detect and thus increase the chances of detecting an impact, though this may be reduced by the lower sensitivity at these magnitudes, with LSST only having 0.1 magnitude precision for the dimmest objects \citep{LSSTBook}. Given that LSST will have an accurate sample of 10 billion stars, 5 times more than \textit{Gaia}, a reasonable first order estimation for the number of expected impacts would be 5 times as many impacts we expect from \textit{Gaia}.

% LSST is expected to contain a different range of stars to \textit{Gaia}. Although Both surveys will observe many of the same nearer stars in the Milky Way disk, LSST will observe a substantial number of stars in the galactic halo \citep{LSSTBook}. Halo stars are among some of the first Milky Way stars to form \citep{BinneyTremaine} and are hence unlikely to host giant impacts. % I'm not really confident about this either

\subsubsection{Frequency of giant impacts} \label{sec:frequency}

In order to determine the anticipated number of impacts in a survey, it is also important to estimate the occurrence rate of giant impacts. An accurate estimation poses a considerable challenge as, not only is the occurrence rate of terrestrial planets uncertain, but the frequency of giant impacts in the history of an average terrestrial planet is much debated \citep{Quintana2016}. In the following sections we synthesize estimates of both these occurrence rates.  

\paragraph*{Planet occurrence rate}

Due to observational biases in various methods of exoplanet detection, it is difficult to estimate the typical number of planets per star. A variety of studies have investigated planet occurrence rate using surveys such as TESS and Kepler and attempted to correct for these observational biases, mainly with a focus on the occurrence rate of Earth-like planets around sun-like stars. 

\begin{itemize}
    \item \cite{Burke2015} estimate around 0.77 terrestrial planets per star.
    \item \cite{Dressing2015} estimate 2.5 planets per M dwarf.
    \item \cite{Shabram2020} estimate between 0.49 and 0.77 terrestrial planets per GK star.
    \item \cite{Hsu2020} estimate between 4.2 and 8.4 planets per M dwarf.
    \item \cite{Kunimoto2020} estimate around 0.9 planets with an orbital period less than 100 days and a radius less than 10 $R_{\oplus}$.
\end{itemize}

All studies used Kepler data. As most stars that can host potentially detectable impacts are M dwarfs, the studies focusing on M dwarf planet occurrence rate are most relevant for our study. 

Given the large amount of uncertainty and range of estimates for the planet occurrence rate, the distribution of the number of planets around a given star will be estimated using a triangular distribution in our MC calculations, with a lower limit of 0, an upper limit of 8 (using the largest figure from \cite{Hsu2020}) and a mode of 3.

\paragraph*{Giant impact rate}

Estimating the average number of giant impacts that take place over the lifetime of planetary systems is difficult, given the small number of such events and the difficulty in detecting them. Nevertheless, there is strong evidence that at least one giant impact has occurred in our own solar system, during the formation of our Moon \citep{Hartmann1975, CameronWard1976}, with several other bodies also proposed to have undergone such a collision. Assuming our planetary system is fairly typical, we can thus assume most planets experience at least one giant impact during their formation.

N-body simulations have been used to simulate planetary system formation and can be used to theoretically estimate the expected number of giant impacts per star system. In a series of N-body solar-system formation simulations from \cite{Quintana2016}, the mean number of giant impacts experienced by an Earth-analogue was three, with an Earth-analogue defined as a planet with a mass greater than 0.5 $M_{\oplus}$, which orbits between 0.75 and 1.5 AU. 

In \cite{Quintana2016}, the number of impacts per planet approximately followed a Poisson distribution. As a first order estimate, we hence assume other non Earth-like terrestrial planets and/or bodies in non-solar like systems experience a similar frequency of giant impacts and use the Poisson distribution from \cite{Quintana2016} to estimate the giant impact occurrence rate in our MC calculations.

\paragraph*{Stellar age distribution}

Solar-system analogues have the vast majority of their giant impacts in their first 100 Myr \citep{Quintana2016}, and so we must also consider what proportion of stars are young enough for giant impacts to be likely to occur. As stellar age estimates are poor, we look at all stars and assume that a certain fraction are young enough to host giant impacts. We can make an initial estimate that all stars have a lifetime of 10 Gyr. This gives a percentage of stars younger than 100 Myr to be 1\%, assuming a uniform distribution of age. Stars with a mass less than 1 $M_{\odot}$ have a lifetime longer than this, and in the case of M dwarfs, have lifetimes closer to 1 Tyr. However the Milky Way disk is about 9 Gyr old and the Milky Way Halo is about 13.5 Gyr \citep{delPeloso2005}. Most of the stellar mass of the Milky Way is in the disk \citep{Licquia2015}. Given the capabilities of \textit{Gaia} and LSST, giant impacts are likely to be detected around low mass M and K stars only, hence 10 Gyr is a reasonable estimate of maximum age.

\section{Results} \label{sec:results}
\subsection{Post-impact emission}

\begin{table*}
    \centering
    \begin{tabular}{cc|cc|cc|cc|cc}
        \multicolumn{2}{c|}{Simulation} & \multicolumn{2}{c|}{} & \multicolumn{2}{c|}{$P = 1$ day} & \multicolumn{2}{c|}{$P = 10$ days} & \multicolumn{2}{c}{$P = 100$ days} \\
        Set & \# & $M_{\mathrm{final}}$ / $M_{\oplus}$ & $AM_{\mathrm{final}}$ / $10^{34}$ $\text{kg}\,\text{m}^2\text{s}^{-1}$ & $L_{0}$ / $10^{-3}L_{\odot}$ & $t_{\mathrm{cool}}$ (days) & $L_{0}$ / $10^{-3}L_{\odot}$ & $t_{\mathrm{cool}}$ (days) & $L_{0}$ / $10^{-3}L_{\odot}$ & $t_{\mathrm{cool}}$ (days) \\ \hline
        A & 0 & 0.99 & 7.33 & 0.38 & 1433 & 0.33 & 1723 & 1.13 & 257 \\
         & 1 & 0.20 & 0.50 & 0.05 & 685 & 0.18 & 200 & 4.07 & 6 \\
         & 2 & 0.50 & 2.32 & 0.17 & 1164 & 0.44 & 514 & 2.35 & 72 \\
         & 3 & 0.74 & 4.55 & 0.23 & 1292 & 0.21 & 1500 & 0.86 & 202 \\
         & 4 & 1.49 & 14.40 & 0.99 & 100 & 0.49 & 1384 & 1.03 & 254 \\
         & 5 & 1.99 & 23.23 & 17.69 & 12 & 2.11 & 437 & 18.03 & 70 \\
         & 6 & 2.98 & 45.50 & 41.92 & 11 & 13.08 & 12 & 27.90 & 80 \\
         & 7 & 3.98 & 73.56 & 48.97 & 16 & 3.50 & 92 & 1.93 & 576 \\
        B & 8 & 0.52 & 0.40 & 0.16 & 351 & 0.62 & 102 & 2.78 & 19 \\
         & 9 & 0.59 & 1.60 & 0.08 & 1337 & 0.05 & 1956 & 0.45 & 99 \\
        C & 10 & 0.74 & 3.97 & 0.29 & 893 & 0.50 & 468 & 1.51 & 99 \\
         & 11 & 0.15 & 0.27 & 0.05 & 299 & 0.19 & 91 & 2.21 & 6 \\
         & 12 & 0.37 & 1.26 & 0.09 & 1087 & 0.30 & 313 & 2.12 & 35 \\
         & 13 & 1.49 & 12.51 & 7.71 & 57 & 160.50 & 13 & 3457.85 & 51 \\
         & 14 & 2.95 & 36.91 & 23.62 & 16 & 3.07 & 570 & 8.15 & 162 \\
        D & 15 & 0.98 & 1.41 & 7.70 & 8 & 5.21 & 8 & 7.07 & 31 \\
         & 16 & 0.98 & 2.80 & 1.32 & 28 & 0.30 & 957 & 0.36 & 317 \\
         & 17 & 0.99 & 4.32 & 7.64 & 9 & 5.25 & 10 & 7.69 & 43 \\
         & 18 & 0.99 & 5.84 & 3.07 & 15 & 1.29 & 240 & 5.96 & 80 \\
         & 19 & 0.99 & 8.74 & 0.41 & 1684 & 1.49 & 514 & 15.08 & 68 \\
         & 20 & 0.99 & 10.25 & 0.37 & 1684 & 2.05 & 437 & 23.59 & 28 \\
        E & 21 & 0.20 & 0.10 & 0.05 & 597 & 0.06 & 447 & 3.18 & 12 \\
         & 22 & 0.49 & 0.46 & 0.24 & 453 & 0.20 & 339 & 4.93 & 16 \\
         & 23 & 1.96 & 4.39 & 35.92 & 7 & 47.98 & 3 & 270.83 & 0 \\
         & 24 & 3.91 & 14.01 & 214.79 & 3 & 252.15 & 4 & 2220.06 & 0 \\
    \end{tabular}
    \caption{This table presents the results of our model applied to our SPH simulations, including the mass of the post-impact body ($M_{\mathrm{final}}$), the final angular momentum of the impact ($AM_{\mathrm{final}}$) and the initial impact luminosity ($L_{0}$) and the cooling timescale ($t_{\mathrm{cool}}$) for 3 different values of orbital period, calculated as described in \S\ref{sec:methods}. The initial conditions of the simulations are shown in Table~\ref{tab:simulation_params}.}
    \label{tab:model_results}
\end{table*}

\begin{figure*}
	\includegraphics[width=\linewidth]{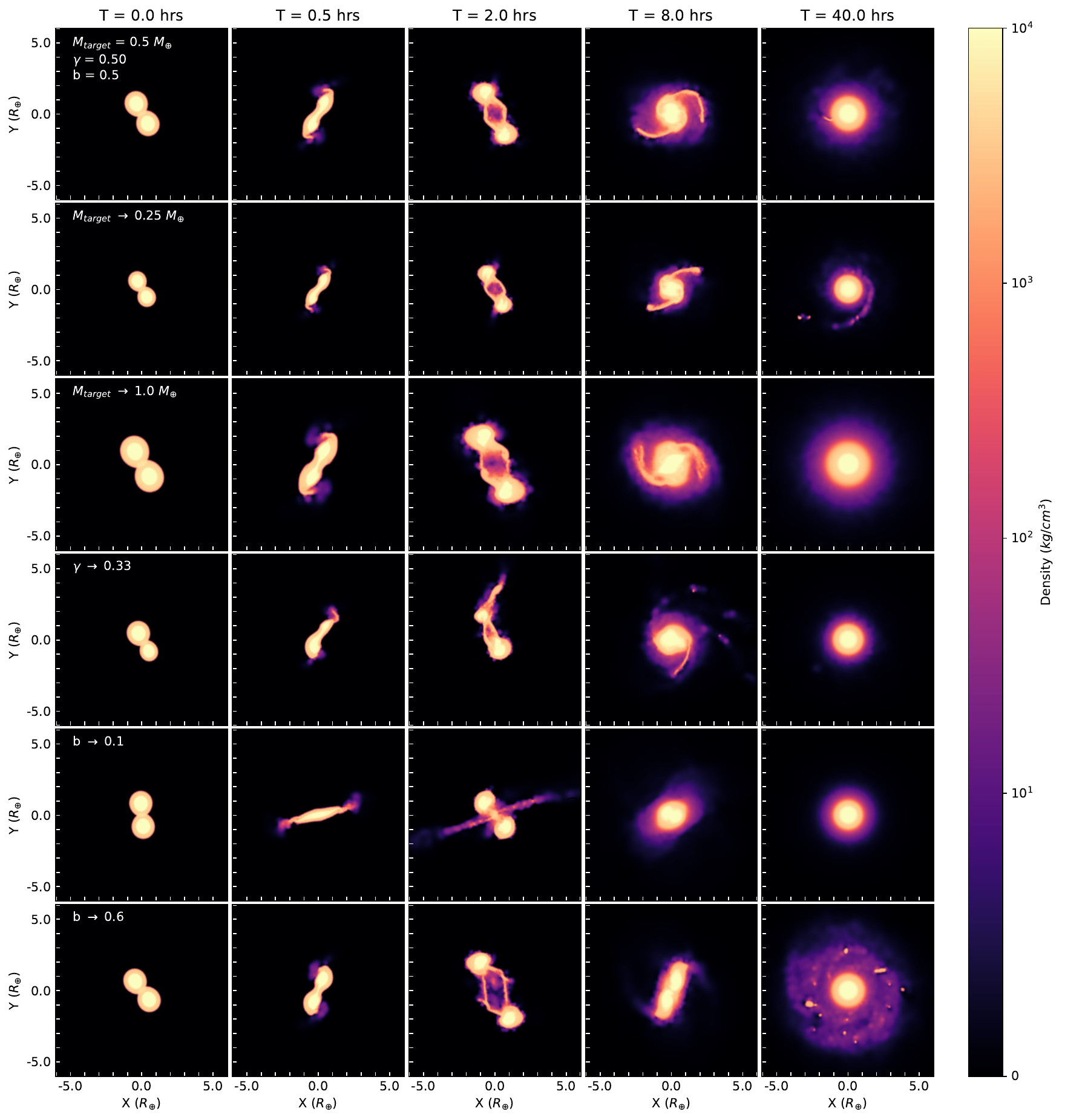}
    \caption{Snapshots of the density in the midplane of the colliding bodies from a selection of impacts with varying impact parameters. The first impact is the baseline scenario, with the rest of the impacts shown having the same initial conditions aside from a single parameter changed. The second and third impacts show the effect that changing the mass has on the impact remnant. The fourth impact shows the effect of changing the impactor-to-target mass ratio. The fifth and sixth impacts demonstrate the effect of altering the impact parameter. }
    \label{fig:impact_plot}
\end{figure*}

\begin{figure*}
    \centering
    \includegraphics[width=\linewidth]{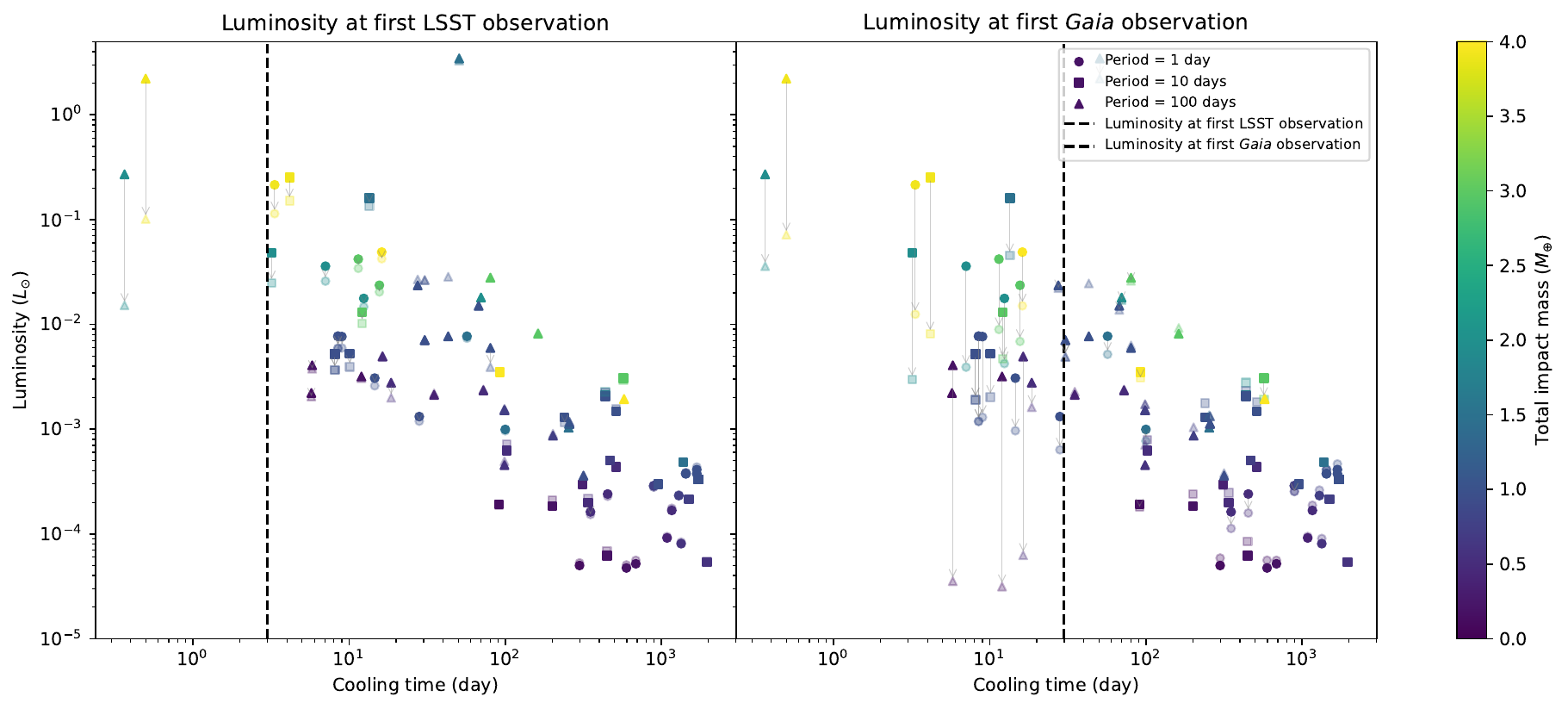}
    \caption{Initial luminosity (luminosity immediately after impact -- an upper bound) and after cadence luminosity (luminosity after cooling for the survey cadence length -- a lower bound), and cooling time of all simulated giant impacts (see also Table \ref{tab:simulation_params}). The total colliding mass is indicated by the colour and the symbols indicate the orbital period of the post-impact body. Higher mass impacts in general have higher luminosities and longer cooling times. Luminosity after cadence is shown by the lighter shaded markers, and shows the minimum luminosity that will be measured by the survey. The average observational cadences of Gaia and LSST are marked on the plot and impacts must be to the right of these lines to be observable by these surveys.}
    \label{fig:results}
\end{figure*}

Density snapshots of different times during a selection of our simulated impacts are shown in Fig.~\ref{fig:impact_plot}, illustrating the effects of changing the impact initial conditions. The physical outcomes of our simulations, for example the final bound mass, are consistent with previous studies \citep{Canup2012, Lock2018}. 
% Increasing the mass of the target and impactor increases the size of the impact remnant, while reducing the mass does the opposite. Reducing the impact parameter makes the impact more violent and results in more mixing, but produces a smaller remnant. Increasing the impact parameter produces a larger remnant. Reducing the mass ratio of impactor to target also leads to a smaller remnant.

Using the model described in \S\ref{sec:methods}, the initial luminosity (defined as the luminosity of the post-impact body calculated at the end of the SWIFT simulation), and cooling timescale (defined as the time taken for the luminosity to half) for each simulated giant impact are calculated and shown in Table~\ref{tab:simulation_params} and Fig~\ref{fig:results}. The simulated giant impacts have luminosities between $10^{-5}$ and 1 $L_{\odot}$. Many of the impacts are about as luminous as an M dwarf and should be sufficiently luminous to be detectable around M or K type stars with \textit{Gaia}. The very brightest impacts cool very fast, falling by an order of magnitude within 30 days. Thus, an appropriate upper bound for observed impact luminosity is 0.1 $L_{\odot}$. The brightest impact was produced by the impact with the highest total mass and there is a rough positive trend between colliding mass and luminosity. However, there are also strong dependencies on other impact parameters and more simulations are needed to fully explore these relationships.  

\begin{figure*}
    \centering
    \includegraphics[width=\linewidth]{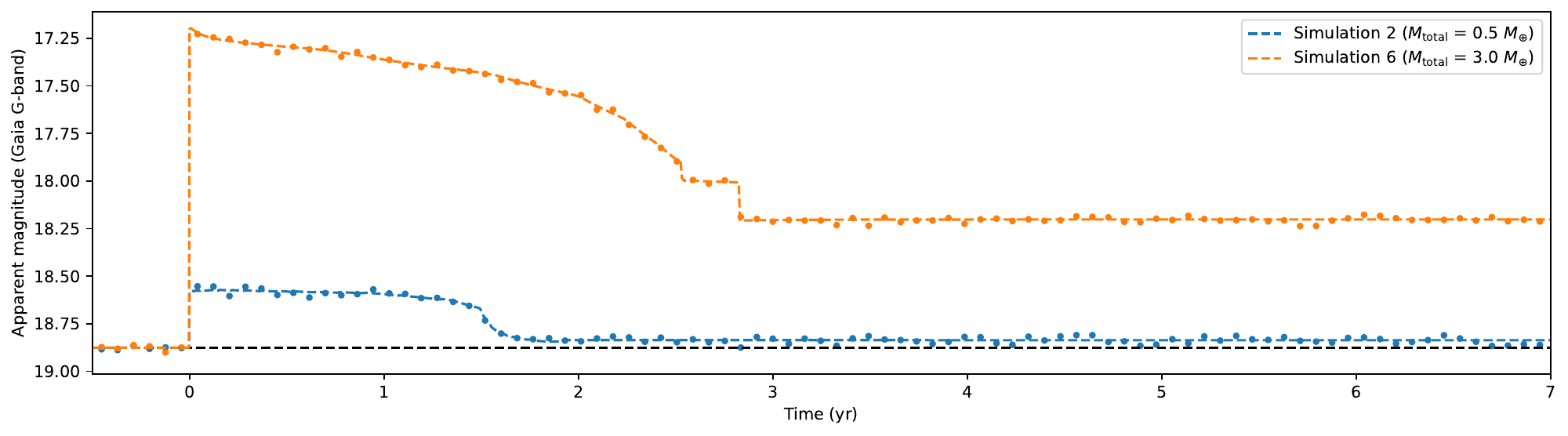}
    \caption{Plot showing simulated \textit{Gaia} epoch photometry observations of two simulated giant impacts (from Table~\ref{tab:simulation_params}), with a sampling rate of 30 days and added random noise calculated from the uncertainties shown in Fig.~\ref{fig:single_flux_error}. The continuous light curve for each giant impact (with no noise) is also shown. The impact occurs at $t=0$. Note the difference in scales on the y-axis.}
    \label{fig:example_light_curves}
\end{figure*}

The estimated cooling timescales are typically between 1 and 2000 days. This is consistent with previous estimates of the full lifetime of Moon-forming post-impact bodies \citep{LockStewart2017, Lock2018,LOCK2020EPSL}.  With \textit{Gaia}'s average cadence being 30 days \citep{GaiaReleaseInfo} and LSST's being 3 days, the visible lifetimes of the majority of the post-impact bodies in our simulation suite are sufficient that the associated change in luminosity of a star could show up in either survey. A selection of example light curves are shown in Fig.~\ref{fig:example_light_curves}. 

The light curve's shape is in three phases. The first phase is the initial spike where the impact occurs, assumed to happen essentially instantaneously. The second phase is where the emission is dominated by the hot vapour atmosphere, which cools and condenses, leading to droplet rainout and steadily decreasing luminosity. The third phase is where the vapour atmosphere has completely condensed and rained on to the post-impact body's liquid surface and the emission is dominated by just the molten remnant.  

The cooling time and initial luminosity are relatively insensitive to variations in the orbital period: varying the orbital period between 3 and 300 days produced less than an order of magnitude change in cooling time and no change in the initial luminosity. This is due to the fact that the outermost regions of the post-impact body contain very little mass and so have only a modest effect on the overall evolution of the body.

\subsection{Observability of giant impacts}

Brighter impacts with luminosity around $10^{-1}$ $L_{\odot}$ can be detected around approximately 80\% of \textit{Gaia} stars, while dimmer impacts with luminosities around $10^{-3}$ $L_{\odot}$ can only be detected around approximately 1\% of \textit{Gaia} stars. 

\begin{figure}
    \centering
    \includegraphics[width=\linewidth]{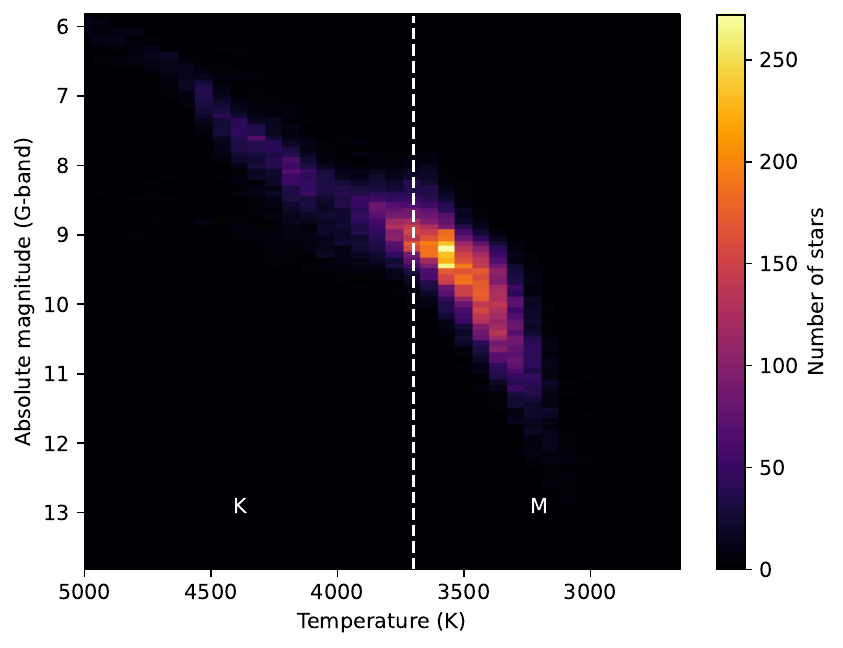}
    \caption{2D histogram showing the distribution of stars in the \textit{Gaia} catalogue which allow a 3-sigma detection of a giant impact on an HR diagram. The vast majority of stars where giant impacts are detectable have an effective temperature below 3700 K and are hence M dwarfs. A small number are K type stars and none are G type or brighter.}\label{fig:HR_detectability}
\end{figure}

Fig.~\ref{fig:HR_detectability} shows the stars in the \textit{Gaia} catalogue that can host potentially detectable giant impacts on an HR diagram, estimating an impact luminosity of $5 \times 10^{-3}$ $L_{\odot}$. The majority of suitable stars have an effective temperature below 3700 K and are thus M dwarfs. Some examples of light curves resulting from a giant impact occurring in orbit of an M dwarf simulated as \textit{Gaia} epoch photometry (as they would appear in \textit{Gaia} DR4), are shown in Fig.~\ref{fig:example_light_curves}.

\subsubsection{Expected number of observed giant impacts}

\begin{figure*}
    \centering
    \includegraphics[width=\linewidth]{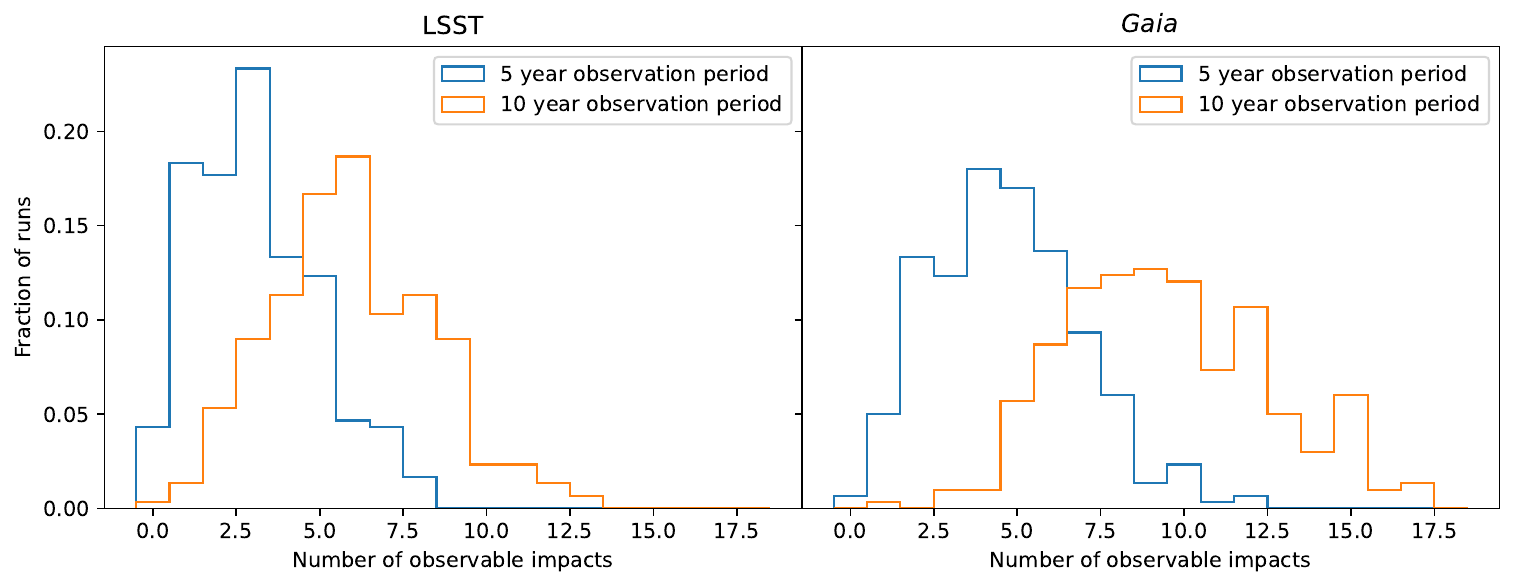}
    \caption{Plot showing the number of post-impact bodies expected to be detectable with \textit{Gaia} and LSST in the \textit{Gaia} catalogue across a 5 and 10 year observing period, calculated using the Monte Carlo method.}
    \label{fig:monte_carlo}
\end{figure*}

The results of the Monte Carlo analysis are shown in Fig.~\ref{fig:monte_carlo}, allowing us to estimate the probabilities of detecting a giant impact with both \textit{Gaia} and LSST.  We assume post-impact bodies have luminosities between $5\times10^{-5} L_{\odot}$ and $1\times 10^{-1} L_{\odot}$, with a mode of $5 \times 10^{-3}$ $L_{\odot}$. For a five year period of observations (corresponding to the duration of observations in \textit{Gaia} DR4), we expect between 0 and 13 impacts to be visible, with an over 90\% chance of observing at least 1 impact. For a ten year period (corresponding to \textit{Gaia} DR5), we expect between 1 and 17 impacts.

For our conservative estimate with LSST (described previously), assuming that LSST observes only \textit{Gaia} stars, we estimate LSST will observe between 0 and 8 impacts in a 5 year period, with an over 90\% chance of observing at least one impact. In a ten year period, we expect between 0 and 14 impacts.

Our more optimistic estimate for LSST assumes that, given LSST will accurately measure 5 times as many stars as \textit{Gaia}, we will detect 5 times as many impacts with LSST. In this case, we expect potentially tens of impacts detected even in a five year period.

\section{Discussions}\label{sec:discussions}

We estimated the luminosities and cooling times for a series of post-impact bodies formed from giant impacts between terrestrial planets simulated with {\small SWIFT}. We found the initial luminosity (40 hours after impact) of a post-impact body was between $5\times10^{-5}$ and $10^{0}$ $L_{\odot}$ and the cooling timescale (time taken for the luminosity to half) was between 1 and 2000 days. This was sufficiently bright and long lasting for many post-impact bodies to be observed around M and some K type stars with \textit{Gaia}. Estimating giant impact occurrence rates, we expect between 0 and 5 impacts in \textit{Gaia} DR4 and between 0 and 6 in DR5, with at least a comparable number detected by LSST.

In this section, we discuss potential limitations to our method of estimating the luminosity and cooling times of these post-impact bodies. We will also consider how a wider range of planet compositions and masses may affect our results. We will discuss potential follow up observations to confirm the presence of a post-impact body and the implication such a detection would have in the field of planetary science.

\subsection{Limitations of luminosity calculations}

\subsubsection{Heat redistribution within post-impact bodies}
\label{sec:discussion:cooling}

Estimating the cooling rate of a post-impact body is challenging, because of the many complex physical and chemical processes present that govern the evolution of the post-impact body \citep[for a more in-depth discussion see][]{Lock2018, Lock2020SSR}. Energy loss from the system is dominated by radiation from the photosphere, but how energy is transported to the photosphere in the multiphase, multicomponent system is uncertain \citep{Lock2018}. The problem is further complicated by the fact that changes to the mass distribution leads to changes in both gravitational potential and kinetic energy \citep{LOCK2020EPSL,Carter2020}. 

In light of this complexity, we have taken a simplified approach to cooling and we now discuss the implications of this for our results.
Our simplified model prescribes that regions below a certain pressure limit lose internal energy equally at each time step, effectively assuming that these outer regions exchange heat efficiently on the timescale of initial cooling. Regions at higher pressures are assumed to not exchange heat with the outer regions, due to thermal stratification or suppression of convection by rotation \cite{Lock2020SSR}, and do not cool during the period of evolution we model here. Consequently, a larger pressure limit allows more energy from the interior to be transferred outwards, buffering the cooling of the outer layers. Conversely, a smaller pressure limit allows the outer regions to cool more quickly.

We took our default pressure limit as $10^9$~Pa, motivated by considering some simple limits to the dynamics of the structure.
\cite{Lock2018} argued that mixing of the outer portion of a post-impact body early in its evolution is dominated by the condensation, infall, and re-vapourization of droplets in the outer regions of the post-impact body. Such infall would force convection and driving a return flow of hotter material to maintain the hydrostatic structure. If droplets (or multi-phase downwellings) were able to fall rapidly compared to thermal equilibration, they would fall isentropically and reach a neutral buoyancy level before vapourizing. In this case, all sections of the structure with an entropy greater than the critical-point for vapourization would be mixed. The exact pressure value this corresponds to is on the order $10^8$--$10^{10}$~Pa depending on the exact impact, with the mixed region encompassing tens of percent of the silicate fraction of the body \citep[see Figure~15A in ][]{Lock2018}. An alternative, more conservative, estimate would be to assume that droplets were heated efficiently as they fell, with their entropies buffered by the latent heat of vapourization. In this case, the maximum pressure droplets would fall to be would be the pressure of the critical point which is on the order of $10^8$~Pa for relevant silicates \citep{Caracas2023,Kraus2012,Townsend2019,ANEOS2019,Xu2025}. An argument could be made that only material in the disk-like region of the structure would be able to mix with itself, with the boundary between the corotating and disk-like regions being a significant barrier to return flow \citep{Melosh2014}. The pressure of this dynamic transition is much more variable between impacts, but for the more energetic impacts that are more likely to be observed it is typically on the order of $10^9$~Pa. Considering these different arguments we conclude that a pressure limit of $10^9$~Pa for mixing of the outer regions of the post-impact body is a reasonable order of magnitude estimate.

\begin{figure}
    \centering
    \includegraphics[width=\linewidth]{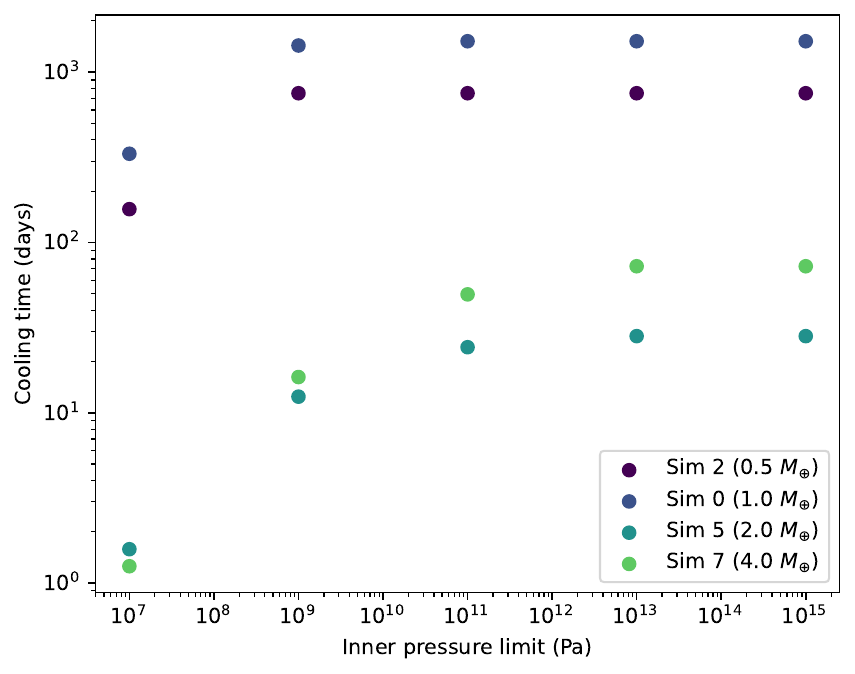}
    \caption{Plot showing how altering the pressure limit changes the post-impact body cooling time for four simulations. At pressure limits above $10^{13}$ Pa, all of the energy of the post-impact body is radiated away. Lower pressure limits lead to a much smaller cooling time. }
    \label{fig:pressure_plot}
\end{figure}

Fig.~\ref{fig:pressure_plot} shows how changing this pressure limit affects the cooling time for a series of impacts of differing total mass (colors). At pressure limits higher than $\sim10^{13}$ Pa, all the mass of the body is included in the mixed region and the entire internal energy budget of the post-impact body participates in the early stages of cooling. This is likely an unphysical scenario, but provides an upper bound for the cooling time. It is unlikely that heat from the core of a planet sized body can escape on such short timescales, especially given that the core of our own planet is still extremely hot, even after nearly 5 Gyr. Reducing the limit from $10^{11}$ to $10^{9}$~Pa  or from $10^{9}$ to $10^{7}$~Pa, the cooling timescale reduces by an order of magnitude. Therefore, if we have overestimated the depth of mixing it is possible that many lower mass impacts would cool more rapidly than we have predicted and so could be potentially unobservable. However, a key process our cooling model neglects is the additional kinetic and gravitational potential energy release during cooling, that would help buffer the cooling of the post-impact body and offset the effect of a lower cooling mass. We therefore consider our estimates of the cooling and emission from post-impact bodies to be a reasonable first approximation, but much work is needed to better understand the cooling and evolution of post-impact bodies.

\subsubsection{Droplet physics}\label{sec:droplet_discussions}

\begin{figure*} % this will stay here for now
	 \includegraphics[width=\linewidth]{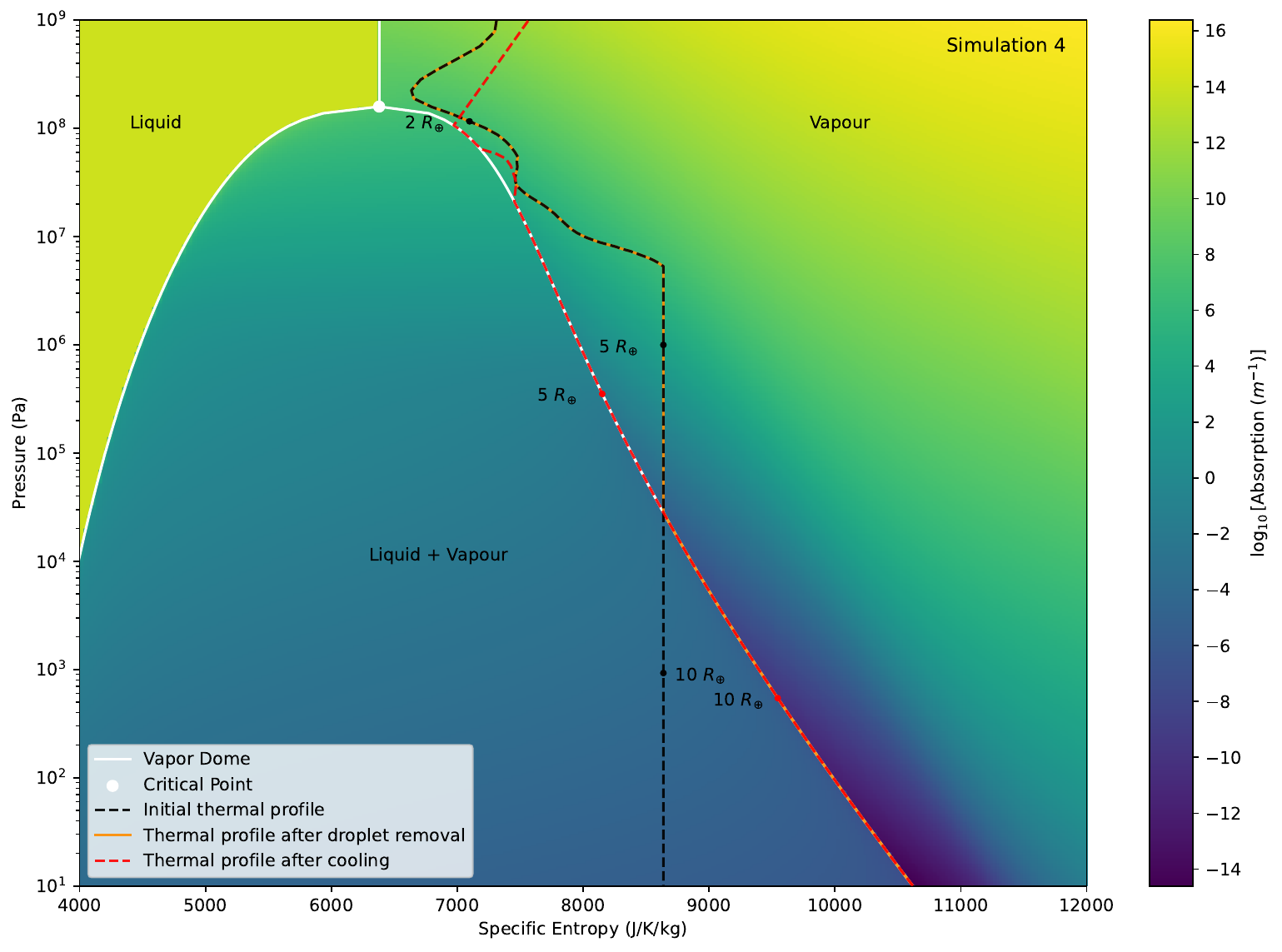}
    \caption{Phase diagram show how the absorption coefficient of silicate changes with specific entropy and pressure, calculated using the forsterite equation of state \citep{ANEOS2019} and the optical absorption model \citep{Kraus2012}, taking into account droplet absorption. The silicate vapour becomes optically thin at the length scales of a post-impact body at pressures below around $10^{4}$ Pa in the vapour phase. The thermal profile, including the model extrapolation, of Simulation 0 is plotted on the phase diagram. The plot shows entropy-pressure profile changing with droplet removal and cooling.}
    \label{fig:phase_diagram}
\end{figure*}

As well as effecting the redistribution of mass, heat, and angular momentum, droplets provide a significant contribution to the absorption of condensing silicate vapour. Fig.~\ref{fig:phase_diagram} shows the photon absorption across the forsterite phase diagram, with the region inside the vapour dome showing significantly higher absorption than low pressure vapour. Given the uncertainty in droplet dynamics, it is difficult to determine what fraction of droplets could remain in the outer regions of post-impact bodies to affect the emission. However, we can provide an estimate of the contribution that condensing droplets have on the opacity of a post-impact body, and hence luminosity and cooling time, by considering two extreme end-member cases. The first case is one with significant droplet infall (the droplet infall model presented in \S\ref{sec:methods} and used in the results presented elsewhere in this work), the second with no droplet infall. In neither case do we consider transport by droplets, or the corresponding response of the vapour.

\begin{figure}
    \centering
    \includegraphics[width=\linewidth]{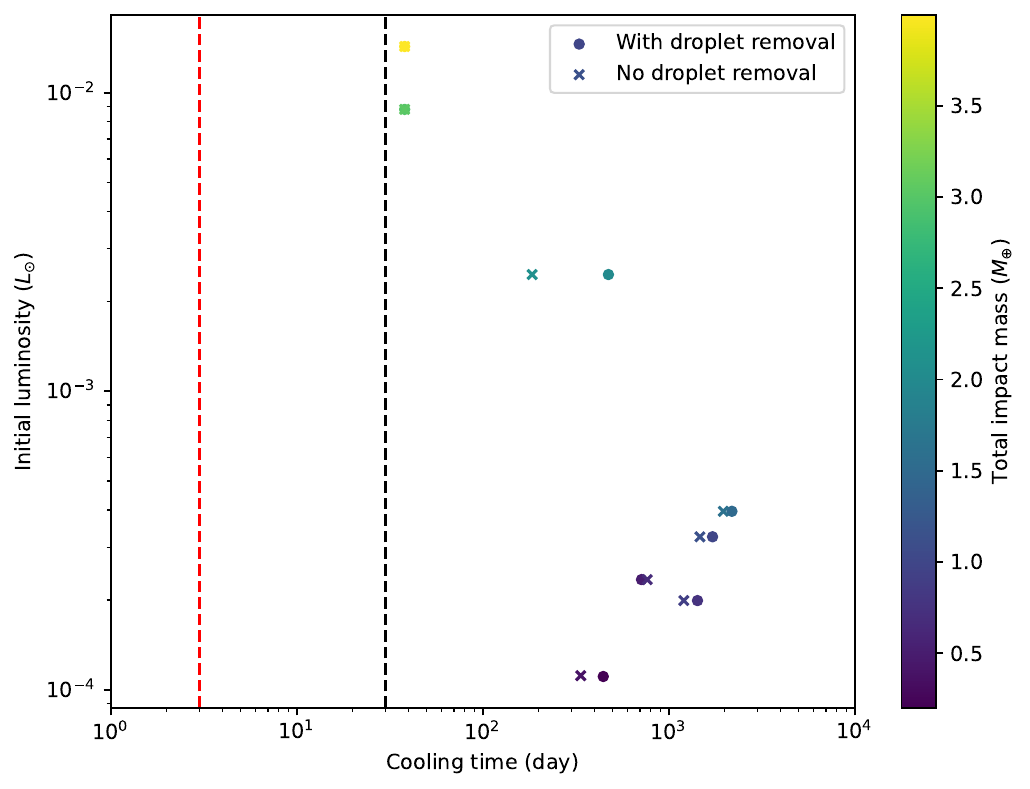}
    \caption{Omitting droplet removal has a minimal effect on the luminosity but can change the cooling time. However, the change in cooling time is not significant enough to alter the observability of any impact. The initial luminosity cooling timescale for Simulations 0 to 6 with their orbital period set at 10 days, as described in \S\ref{sec:droplet_discussions}.}
    \label{fig:droplets}
\end{figure}

Fig.~\ref{fig:droplets} shows the luminosity and cooling time for an example set of impacts in these two cases. The lack of droplet infall does not significantly change the calculated initial luminosity. The cooling time is more affected by the lack of droplet infall, with the cooling time getting shorter in some cases and longer in others. However, this change is not sufficient to change the detectability as the cooling timescales remain above 30 days (the average time between observations with \textit{Gaia}). 

\subsubsection{The effect of unbound material}

Here, we have considered only the luminosity and cooling of material that is bound to the post-impact remnant. However, there could also be a significant emission from material that becomes unbound in the collision. This material expands over a large volume, and can be substantially vapourized and so radiating at high temperatures. The ejecta could therefore add significantly to the luminosity from the system that hosted the impact for a short period after the impact. The ejected material quickly cools and so this contribution to the luminosity in the optical is only likely to last a few days and so only recorded in, at most, one observation from \textit{Gaia} or LSST. Nevertheless, if the impact occurs sufficiently close to the star that the ejecta is passively heated by stellar radiation, it could still produce a significant excess flux in the infrared as seen in various systems \citep[e.g.,][]{Su2019,Su2022}. The existence of warm ejecta may aid in confirming the identity of a source as a giant impact and even a single \textit{Gaia} or LSST detection in such a case would still provide significant extra information for understanding the nature and evolution of the impact (Section~\ref{sec:discussion:confirming}).

The amount of material that becomes unbound in the impact would also have a significant effect on the long-term luminosity and cooling timescales of the post-impact body. As our impact simulations were conducted in the absence of a star, they give an upper bound on the bound mass. In reality, material from the post-impact body outside the Hill sphere will become unbound \citep[see][for a study of this effect in circum-planetary impacts]{Rufu2018}, reducing the size of the post-impact body and potentially changing both its luminosity and cooling time.

\begin{figure}
    \centering
    \includegraphics[width=\linewidth]{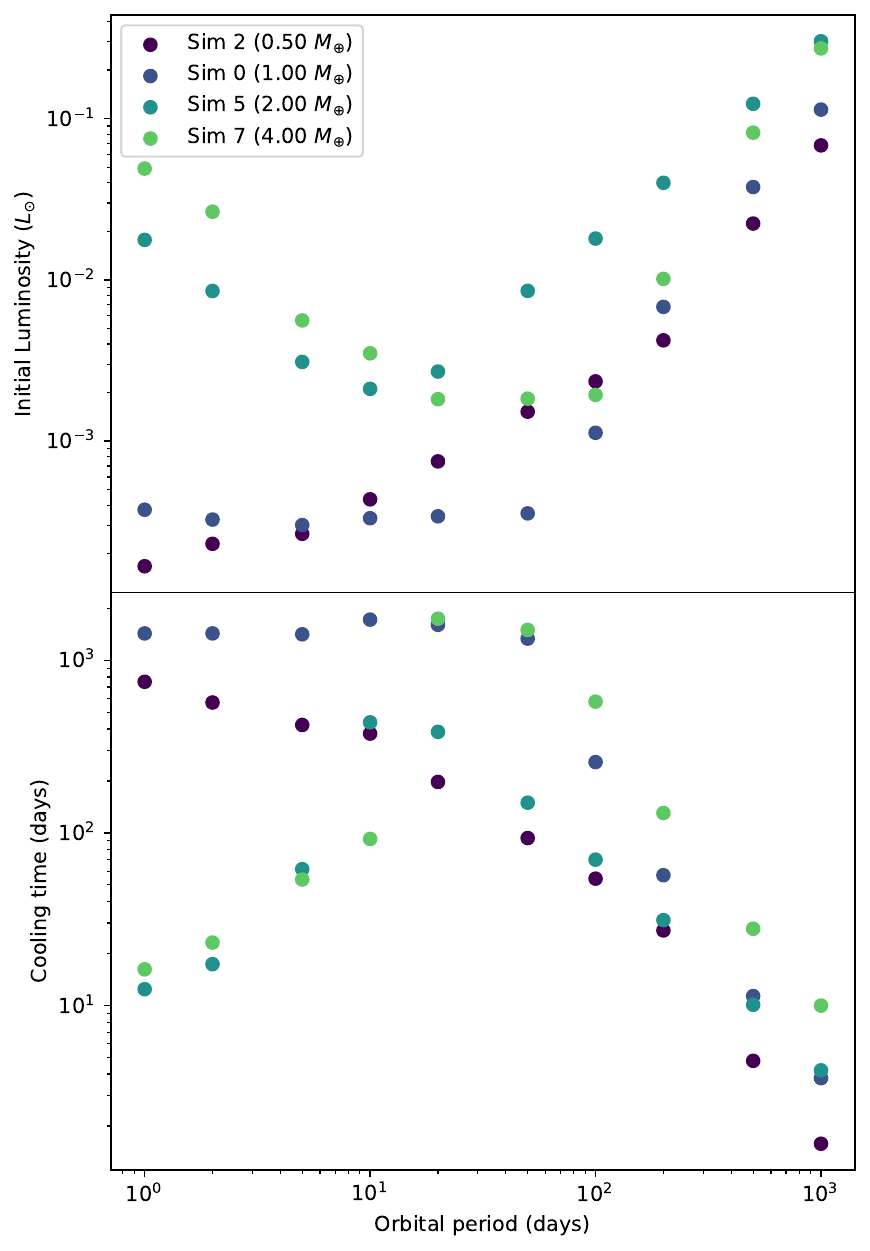}
    \caption{The effect of altering the orbital period of the post-impact body on the initial luminosity and cooling time for a range of simulations.}
    \label{fig:hill_plot}
\end{figure}

To investigate the effect of the orbital location of a giant impact, we calculated the effect of altering the orbital period for a group of simulations with differing total masses, with the results shown by Fig.~\ref{fig:hill_plot}. To account for the effect of the Hill sphere we have post hoc assumed that only mass inside the Hill sphere contributes to the emission from the post-impact body. 

All of the simulations show increasing luminosity with higher orbital periods (larger Hill spheres) due to more of the hot vapourised silicate envelope being retained and an increased emitting area. This leads to a shorter cooling time. The higher mass simulations show increased luminosity with shorter orbital periods (smaller Hill spheres) due to the cooler exterior of the cloud being lost, exposing the hotter interior, which is large enough to reach the Hill sphere. The lower mass simulations luminosities plateau at lower orbital periods because the cloud is sufficiently compact that the photosphere lies well within the Hill sphere even at short orbital periods. The Hill sphere is outside the hot interior, so neither the emitting area nor the photosphere temperature changes significantly with period -- the photosphere surface remains the same regardless of the Hill sphere size. Only at longer periods, when the isentropic extrapolation extends the atmosphere far enough outward, does the photosphere radius grow appreciably and the luminosity begin to rise.

To more accurately determine the material that would make up the bound post-impact body would require running SPH simulations with an external gravitational field and a larger number of particles. Just considering the bound mass from such simulations would remove the need to artificially cut off the optical depth integration at the Hill sphere. This would, however, increase the parameter space and render the study much more computationally intensive, as a variety of different orbital parameters would have to be simulated for each impact.

% \subsubsection{Simulation limitations}

% We are limited by the resolution of our simulations (which were constrained by the computing time available). A much higher resolution simulation, with a much larger number of particles, would reduce or even remove the need for the extrapolation, as we would be able to resolve the outer regions where the post-impact body goes from being optically thick to thin. These regions are crucial for estimating the luminosity of the structure. A higher resolution simulation would allow us to confirm and refine our extrapolation model, but it is currently too computationally expensive to run even one sufficiently high resolution simulation.  

% The discussion on the Hill sphere of the post-impact body demonstrates that the gravitational field from the host star can play a significant role in its size, especially for lower mass stars. High resolution simulations that can investigate the vapour at the edges of the post-impact body should take into account that whole system is orbiting the star and hence add accelerations from the star's gravity and the centrifugal force.

\subsection{Confirming identification of giant impact}
\label{sec:discussion:confirming}

Our cooling estimations show that a post-impact body should be visible in a star's light curve as a very sudden brightening (happening on the timescale of a day) followed by a gradual dimming (with a timescale of a few months to years (see Fig.~\ref{fig:example_light_curves}). This is a distinct aperiodic signal that should be identifiable, at least in theory. Nevertheless, there may be a few challenges in confirming the presence of a giant impact from such a light curve.

Stellar variability may cause some difficulties in observing and identifying giant impacts. As giant impacts cause changes in brightness from 1-10\%, a similar magnitude of stellar variability could mask the increase in overall brightness due to a giant impact within the star's normal periodic variations. However, the increase and decrease in luminosity associated with giant impacts are on different timescales (typically several months) to most stellar variability (order days). This difference in timescales should make giant impacts identifiable from stellar variability given observations over a long enough time period. Furthermore, M dwarfs typically flare in the ultraviolet, while post-impact bodies have all of their emission in the infrared, providing a second method of identification. The infrared dominated nature of the giant impact signal may be identifiable with \textit{Gaia} BP/RP epoch spectra and LSST's \textit{ugrizy} filter set \citep{LSSTBook}.  

Another major source of variability in star systems that may pose difficulties for identifying giant impact signals is dust. Both giant impacts \citep{Su2022, Kenworthy2023, Watt2021, Watt2024MNRAS} and processes in debris disks (such as collisional cascades) \citep{Wyatt2008} could lead to infrared variability and an infrared excess in the star's spectra caused by warm dust. However, debris disks emit at longer wavelengths than terrestrial post-impact bodies. As rock-dominated post-impact bodies radiate at around 2000-3000 K (much hotter than any dust in a debris disk which very quickly reaches equilibrium temperature from the star) its IR excess is very near the star’s own blackbody spectra. As warm dust is much cooler than its host star (typically around $10^{2}$ K, but dependent on semi-major axis), its emission is clearly distinct from the star’s blackbody emission. Only impacts very close to their host stars would produce dust that could be passively heated to temperatures similar to those of rocky post-impact bodies. Hence, any process that could produce a large increase in dust is unlikely to be confused with a giant impact, as it would lack a near infrared excess caused by the much hotter post-impact body. 

Dust may also allow giant impacts to be distinguished from stellar variability. Stellar variability will not cause a sudden infrared excess, while warm dust ejected from a giant impact will. If any giant impact candidates are found in \textit{Gaia} or LSST, it would be useful to observe the candidate star system in the mid and far infrared to measure any increase in dust emission that may come with such an impact. This dust is likely to linger in the system for much longer than the post-impact body's cooling time, causing long-term infrared variability around the star. The combination of a single, or few, \textit{Gaia} or LSST observations with observations of IR dust emission may also allow for the identification of post-impact bodies with lifetimes too short to be confirmed by \textit{Gaia} or LSST alone.

\subsection{Observability of the planet zoo}

The analysis presented here only examines impacts involving rocky planets with total colliding masses up to 4 $M_{\oplus}$, and mostly impacts with a total colliding mass of 1 $M_{\oplus}$.  Collisions involving more massive super-Earths (with masses up to 10 $M_{\oplus}$), sub-Neptunes, Neptunes and gas-giant planets have not been considered. Such planets make up a large number of known exoplanets with super-Earths thought to be the most common type of terrestrial planet \citep{Neil2020}. This wider range of planet masses and compositions will affect our estimations of the observability of post-impact bodies. We observed a rough positive relationship between total colliding mass and luminosity, and so would expect impacts involving these higher mass planets should to brighter. Here, to quantify this further, we estimate a scaling relationship between the energy of the impact, luminosity of the impact, and cooling timescales.

The energy participating in an impact is approximately the kinetic energy of colliding planets. We assume that planet density is constant, with $M \propto R^{3}$ (where $M$ and $R$ are the mass and radius of a planet), which is valid for rocky planets. The impact velocities ($v$) are typically on the order of the mutual escape velocity ($v_{\mathrm{esc}}$), which scales with $\sqrt{\frac{M}{R}} \propto M^{1/3}$. Hence the impact energy ($E$) scales as

\begin{equation}
    E \approx \frac{1}{2}Mv^{2}  \propto  M^{5/3}.
\end{equation}

The luminosity ($L$) of the post-impact disk (with radius $r$) is proportional to its emitting area, which we have assumed in this study scales as $r^{3}$. Hence we can derive a scaling relationship for the luminosity assuming that the average density of the post-impact body remains similar and so mass scales with the disk volume ($V \approx r^{3}$).

\begin{equation}
    L \propto r^{3} \propto M
\end{equation}

\noindent An alternative, more conservative, estimate for this relationship could be made by assuming that the scale height of the post-impact disk is approximately the size of the central planet. In this case, the emitting area scales as $r^2$, $V \approx r^{2}R$, and  

\begin{equation}
    L \propto r^{2} \propto M^{2/3} \; .
\end{equation}

\noindent We thus also get a scaling relationship for the cooling timescale ($t_{\mathrm{cool}}$).

\begin{equation}
    t_{cool} \approx \frac{E}{L} \propto M^{2/3}
\end{equation}

\noindent or 
\begin{equation}
    t_{\rm cool} \approx \frac{E}{L} \propto M
\end{equation}
\noindent in the more conservative case.

These scaling relations suggest that impacts involving larger mass planets should be brighter and produce longer lasting signals. A larger variety of stars may be able to host potentially visible giant impacts between more massive bodies and the estimated number of giant impacts given in this work may be an approximate lower bound.

The composition of the planets involved in a giant impact will affect the emission from the post-impact body. This work has only considered rocky planets, like Earth and other rocky planets in our Solar System. Collisions between hydrogen, water or other volatile dominated planet impacts will likely produce larger more highly vapourised post-impact bodies due to the lower latent heat of vapourisation and typically lower densities of lower mean molecular weight gases. The lower vapourisation temperatures of volatile bodies may also result in lower emission temperatures. Assuming a black body emitting temperature 
\begin{equation}
    L \propto T^4
\end{equation}
and 
\begin{equation}
    t_{\rm cool} \propto T^{\frac{1}{4}}
\end{equation}
\noindent so we expect the emission temperature to have a dominant effect on observability.

A particularly illuminating example is the potential giant impact remnant discovered by \cite{Kenworthy2023}, around ASASSN-21qj, a Sun-like star. The proposed impact remnant had a blackbody temperature of around 1000 K, much lower than the post-impact bodies studied in this work (which had temperatures of around 2300-3000 K) and previous works \citep{Lock2018}, but had a radius of 750 $R_{\oplus}$, much larger than those produced by terrestrial giant impacts. As a result, the object was much brighter, causing a 0.8 magnitude increase in the NEOWISE infrared W2 band. The large size and lower emission temperature of the remnant  was suggested to be due to the large mass of the colliding bodies (on the order of 10 $M_{\oplus}$) and a larger amount of volatiles in the planet's composition. The presence of one impact discovered in the ASASSN survey gives a potential constraint on giant impact frequency. In the latest release of ASASSN data, gathered over nearly 10 years, there are 98 million stellar sources in its catalogue, with limiting magnitudes of 17.5 and 18.5 in the \textit{V} and \textit{g} bands respectively \citep{Hart2023}. With one impact discovered in nearly 10 years, we can assume surveys with better capabilities, such as \textit{Gaia}, should discover more than one impact in a similar time frame.

To fully explore the parameter space of planetary impacts and get a more complete picture of the probability of observing a post-impact body, a wider range of impacts should be simulated and modelled. Impacts involving planets with more volatile rich compositions and higher masses is a worthwhile area of exploration, given that these planets are common and their post-impact bodies are easier to observe. Incorporating different materials into the optical depth calculations would be necessary to model such post-impact bodies. 

\subsection{Implications of detecting one or more giant impacts}

The number of giant impacts that will be detected by the \textit{Gaia} and LSST surveys strongly depends on the occurrence rates of planets and giant impacts in the galaxy. Presently, these occurrence rates remain highly uncertain and detection of giant impact remnants would allow bounds to be put on these parameters.

In particular, the role that collisions play in planet formation (particularly of the terrestrial planets) is heavily debated. Classic planet formation theory holds that planets grow by the collision of increasingly large bodies, culminating in a period of giant impacts between planetary embryos \citep{Safronov1972,Chambers2010}. In our solar system, growth from Mars-mass embryos to the final planets would have been largely achieved by giant impacts with each final planet experiencing several giant impacts during their formation \citep{Quintana2016}. More recently, it has been suggested that terrestrial planets could grow to a significant fraction of their final mass through the process of pebble accretion where accumulation of smaller ($<10$~m) particles is accelerated by gas drag in the atmospheric envelopes of proto-planets \cite{Johansen2017,Johansen2021}. In such models, giant impacts are a relatively infrequent event and more a side product of planet formation than the main driver for growth. If pebble accretion were the dominant mode of terrestrial planet formation, the number of giant impacts that would be occurring in the galaxy would be a lot lower than has traditionally been considered.

Constraining the occurrence rate of giant impacts would give valuable insights into the later phases of planetary system formation. Detections of giant impact remnant would provide empirical evidence supporting the involvement of giant impacts during the later stages of planetary system formation, with the frequency of detection placing a constraint on the dominance of the their role in planet formation. Follow up studies of impact remnants may also allow us to investigate the composition of planets that are undergoing giant impacts \cite[e.g.,][]{Kenworthy2023}. A null result, i.e., no detections of giant impacts, would be equally interesting and suggest that pebble accretion was the more dominant driver for planetary growth.

\begin{comment}
These instruments will mainly detect events involving large single impacts that produce long-lasting post-impact bodies, rather than many smaller impacts. We therefore anticipate favouring detection of synestia-forming impacts \citep{LockStewart2017} as such impacts produce remnants with sufficient duration to be detected with the cadence of \textit{Gaia} or LSST. 
\end{comment}

\section{Conclusions}\label{sec:conclusions}

Giant impacts are likely common occurrences in the history of terrestrial planets. This work assesses the potential to observe the aftermath of giant impacts, focusing on \textit{Gaia} and LSST. We present a method to trace the initial luminosity and long term evolution of remnants formed by giant impacts simulated with SPH.

We find that luminosities after a giant impact are between $5\times10^{-5}$ to $10^{-1}$ $L_{\odot}$, and so potentially as bright as an M dwarf. The long term evolution of the post-impact body was estimated, in order to give a lower bound estimate for the cooling time. We found that post-impact bodies had cooling timescales, defined as the time required for the luminosity to half, between 1 and 2000 days. Based on our calculated luminosities and lifetimes, it was found that around 5\% stars in the \textit{Gaia} catalogue could host a potentially detectable giant impact and we predict between 0 and 13 impacts to produce observable signatures in \textit{Gaia} epoch photometry and at least a comparable number detected by LSST (though potentially 5 times more in the best case). The vast majority of these identified impacts would be around M or K dwarfs. 

This work predicts that a very sudden brightening (on the timescale of a day), followed by a gradual dimming over tens to hundreds of days, would be a strong indicator of a giant impact. Such events could then be followed up by more detailed observations, perhaps in the infrared to search for dust produced by the impact and the spectral signatures of planetary composition. With large datasets from large scale surveys imminent, we may soon be able to gain direct insights into a key process in planet formation.

\section*{Acknowledgements}

PT acknowledges support from the Leverhulme Centre for Life in the Universe and the UK Science and Technology Funding Council (grant ST/V000888/1). SJL acknowledges support from the UK Natural Environment Research Council (grant NE/V014129/1). A.B. acknowledges support of a Royal Society University Research Fellowship, URF\textbackslash\textbackslash R1\textbackslash\textbackslash 211421.

The research in this paper made use of the {\small SWIFT} open-source simulation code
(http://www.swiftsim.com, Schaller et al. 2018) version 0.9.0. 
This work was performed using resources provided by the Cambridge Service for Data Driven Discovery (CSD3) operated by the University of Cambridge Research Computing Service (www.csd3.cam.ac.uk), provided by Dell EMC and Intel using Tier-2 funding from the Engineering and Physical Sciences Research Council (capital grant EP/T022159/1), and DiRAC funding from the Science and Technology Facilities Council (www.dirac.ac.uk).

This work also made use of the computational facilities of the Advanced Computing Research Centre, University of Bristol. 

Software used: Matplotlib \citep{matplotlib}, NumPy \citep{numpy}, SciPy \citep{scipy}, SWIFTsimIO \citep{swiftsimio}, unyt \citep{unyt}, WoMa \citep{woma}, Astropy \citep{Astropy2013, astropy2018, Astropy2022}, Astroquery \citep{Astroquery}, \textit{Gaia}dr3 Bcg \citep{BolometricCorrection}

\section*{Data Availability}

Most of the software used in this work is publicly available.
The scripts, input files, and data specific to this work will be made available upon acceptance on GitHub and archived on Zenodo.



\bibliographystyle{mnras}
\bibliography{biblio}

%%%%%%%%%%%%%%%%%%%%%%%%%%%%%%%%%%%%%%%%%%%%%%%%%%

%These appendix plot are probably not needed for the paper's results or conclusion
%They may be helpful in explaining the method

%\appendix

%\section{Extra plots}

%\begin{figure}
%    \centering
%    \includegraphics[width=\linewidth]{figures/rotation.pdf}
%    \caption{Histogram showing the distribution of SPH particle distance to the central rotation axis and rotational velocity for Simulation 2. Within $10^{7}$ m, all the particles are co-rotating as a solid planet. Outside this co-rotation limit, the particles orbit at sub-Keplerian velocity, due to gas pressure providing extra support against gravity. }
%    \label{fig:rotation_plot}
%\end{figure}

% Don't change these lines
\bsp	% typesetting comment
\label{lastpage}
\end{document}